\documentclass[letterpaper,usenatbib,usegraphicx]{mn2e}
\usepackage{times}
\usepackage{amssymb}
\usepackage{amsfonts}
\usepackage{hyperref}
\usepackage{bm}
\usepackage{color}

\newcommand {\ie} {{\it i.e.}}

\newcommand {\eg} {{\it e.g.}}
\newcommand {\ea} {{\it et~al.}}

\newcommand {\be} {\begin{equation}}
\newcommand {\ee} {\end{equation}}
\newcommand {\bea} {\begin{eqnarray}}
\newcommand {\eea} {\end{eqnarray}}

\newcommand{\der}[2]{\ensuremath{\frac{{\rm d} #1}{{\rm d} #2}}}

\newcommand{\pder}[2]{\ensuremath{\frac{\partial #1}{\partial #2}}}

\newcommand{\jet}{_{\rm jet}}
\newcommand{\obs}{_{\rm obs}}

\title[Radiative Properties of Minijets]{Radiative Properties of Reconnection-Powered Minijets in Blazars}

\author[Nalewajko \ea]
{
Krzysztof Nalewajko,$^1$\thanks{E-mail: knalew@camk.edu.pl}
Dimitrios~Giannios,$^2$
Mitchell~C.~Begelman,$^{3,4}$
\newauthor
Dmitri~A.~Uzdensky$^4$
and Marek~Sikora$^1$\\
$^1$Nicolaus Copernicus Astronomical Center, Bartycka 18, 00-716 Warsaw, Poland\\
$^2$Department of Astrophysical Sciences, Peyton Hall, Princeton University, Princeton, NJ 08544, USA\\
$^3$Joint Institute for Laboratory Astrophysics, University of Colorado, Boulder, CO 80309, USA\\
$^4$Center for Integrated Plasma Studies, Department of Physics, University of Colorado, Boulder, CO 80309, USA
}

\begin{document}

\maketitle

\begin{abstract}
We construct a numerical model of emission from minijets, localized flows driven by magnetic reconnection inside Poynting-flux-dominated jets proposed to explain the ultrafast variability of blazars. The geometrical structure of the model consists of two wedge-like regions of relativistically flowing gas, separated by a stationary shock. The dynamics is based on solutions of relativistic magnetic reconnection with a guide field from Lyubarsky (2005). Electron distributions in each region are chosen to the match the pressure and density of the local plasma. Synchrotron emission from both regions is used to calculate Compton scattering, Compton drag and photon-photon opacity effects, with exact treatment of anisotropy and the Klein-Nishina regime. Radiative effects on plasma are taken into account, including the dependence of pressure on electron radiative losses and adiabatic heating of the flow decelerating under Compton drag. The results are applied to the July 2006 flare in the BL Lac object PKS 2155-304, with the aim of matching TeV flux measurements by \emph{H.E.S.S.} with models that satisfy the variability constraints, while keeping X-ray emission below simultaneous \emph{Chandra} observations. We find that models of isolated minijets with a significant guide field overproduce X-ray emission, and that we must take into account the radiative interaction of oppositely-oriented minijets in order to achieve a high enough dominance by Comptonized TeV radiation.  We argue that such interactions are likely to occur in a jet where there is substantial internal reconnection, producing a large number of misaligned minijets. Finally, we show that large jet magnetizations are indeed required to satisfy all observational constraints and that the effective Lorentz factor of the minijet plasma has to be larger than 50, in agreement with earlier one-zone estimates.
\end{abstract}


\section{Introduction}

Relativistic jets produced in Active Galactic Nuclei (AGNs) are thought to be launched as cold and highly-magnetized outflows, the energetics of which is dominated by Poynting flux. Under such conditions shocks are very inefficient, however energy dissipation can be provided by relativistic magnetic reconnection. This idea led to the development of the minijets model, which has been proposed to explain extremely fast variability observed in TeV blazars \citep{2009MNRAS.395L..29G} and radio galaxy M87 \citep{2010MNRAS.402.1649G}. Minijets are perpendicular relativistic flows within a relativistic jet and as such attain very high Lorentz factors $\Gamma>50$ required to circumvent the gamma-ray opacity problem \citep{2008MNRAS.384L..19B}. A number of alternative solutions for this `Lorentz factor crisis` \citep{2006ApJ...640..185H} have been proposed \citep{2007ApJ...671L..29L,2008MNRAS.383.1695S,2008MNRAS.386L..28G,2008MNRAS.390..371K,2008MNRAS.390L..73B,2010arXiv1004.2430L}, considering either significant deceleration of the inner jet or a multi-zone structure of the emitting region.

To our knowledge, there have been no new reports of TeV flare with few-minute timescales, besides the July 2006 outburst of PKS 2155-304 \citep{2007ApJ...664L..71A} and the June-July 2005 events in Mrk 501 \citep{2007ApJ...669..862A}. However, there are more details known about the former event. \cite{2010arXiv1005.3702H} showed that the high-activity state of PKS 2155-304 lasted for only a few days and consisted of many closely following 5-10 min long flares. This was a truly exceptional event, as during the remainder of the years 2005-2007 the source was quiet and remarkably stable. The TeV flux distribution is consistent with a superposition of two log-normal distributions of different spectral behaviour, indicating that the underlying mechanism of the flare is different from the one responsible for the quiescent state. Simultaneous X-ray and optical monitoring \citep{2009A&A...502..749A} indicates strong Compton dominance and an almost cubic correlation between TeV and X-ray fluxes. This indicates that soft emission from the dominant gamma-ray producing region is swamped by radiation produced in a different region. A viable model of extreme TeV flares should account for their apparently very low duty cycle and very high Compton/synchrotron luminosity ratio.

In previous works, the minijet emitting region has been treated as a compact blob propagating through the main jet. Here, we investigate a more physically motivated model. We consider stationary outflows fueled by steady relativistic Petschek reconnection in one or more locations of the main jet. As the fluid is eventually slowed down by a terminal shock separating it from the so-called magnetic islands, two emitting regions that are relativistically boosted with respect to each other form, enhancing the relative importance of the inverse Compton (IC) process. In addition, if many such minijets form in closely aligned pairs with opposite propagation directions \citep{2010MNRAS.402.1649G}, it is plausible that these emitters will interact radiatively, leading to even greater Compton dominance. We study the effects of the presence of a weak guide field, radiative cooling and Compton drag. Given the magnetization parameter required to obtain high enough bulk Lorentz factors, $\sigma\sim 100$, magnetic reconnection will accelerate electrons to random Lorentz factors $\gamma\sim 10^4$, which puts the Klein-Nishina limit for observed photon energy in the TeV band. Thus it is also important to carefully model Klein-Nishina effects on TeV emissivity.

We will introduce a physical scenario leading to the emergence of the minijets in Section \ref{sec_origin}, the geometrical and dynamical structure of a single minijet in Section \ref{sec_struct} and a model for broad-band emission in both analytical and numerical approaches in Section \ref{sec_em}. Our results are presented in Section \ref{sec_res}, followed by a discussion in Section \ref{sec_disc} and a summary in Section \ref{sec_sum}.

\section{The origin of minijets}
\label{sec_origin}

Magnetic fields dominate the energy flux in the inner part of AGN jets \citep{2005ApJ...625...72S} and hence they preserve an order imprinted at the launching region. To launch a relativistic outflow from an accreting black hole system, a strong poloidal magnetic field component is required \citep{2010LNP...794..233S}. It could be either produced locally within an accretion disc by the Parker instability \citep[\eg,][]{1992MNRAS.259..604T} or advected from the galactic environment \citep[\eg,][]{2005ApJ...629..960S,2008ApJ...677..317I,2008ApJ...677.1221R,2009ApJ...707..428B}. Both scenarios can naturally produce separate regions of opposite magnetic polarity. Their topology will be preserved into the innermost region of the accetion disk \citep{2008ApJ...678.1180B}.

\begin{figure}
  \includegraphics[width=\columnwidth]{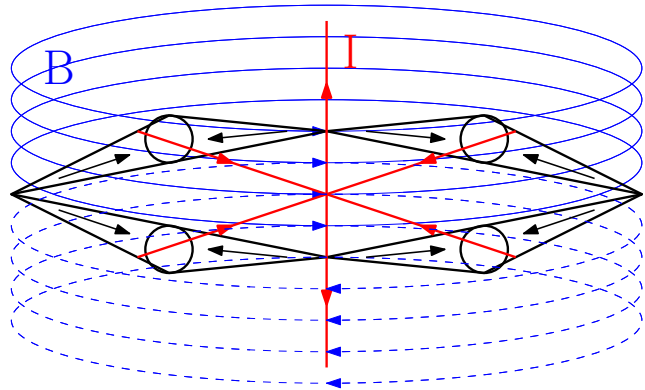}
  \caption{Schematic representation of a toroidal current sheet at the boundary of two magnetic domains of opposite polarity, showing orientations of magnetic fields \emph{(blue)}, currents \emph{(red)} and minijet flows \emph{(black arrows)} from X-points to O-points.}
  \label{fig1}
\end{figure}

In the simplest configuration, two such domains are separated radially in the accretion disk and then consecutively pass through the jet-launching region \citep{1997ApJ...484..628L}. This results in two magnetic domains in the inner jet separated by a current sheet \citep{2003astro.ph..9504S}. As the magnetic fields of either domain are quickly stretched in the toroidal direction by lateral expansion of the jet \citep{1984RvMP...56..255B}, the current sheet can approach a toroidal, disk-like shape. Due to tearing mode instability, it separates along the azimuthal angle into a sequence of X-points and O-points (see Fig. \ref{fig1}). Minijets form as coherent flows from the X-points to the O-points. If the magnetization parameter of the jet is large, the minijets are relativistic in the frame co-moving with the current-sheet. They are directed mainly in the toroidal direction, {\ie} perpendicularily to the jet propagation in the jet rest frame. They are observed as much more relativistic flows than the jet itself \citep{2009MNRAS.395L..29G}.

Alternatively, the minijets could arise without the necessity for magnetic polarity reversal, if the Poynting-flux-dominated jet undergoes kink instabilities \citep[\eg,][]{1993ApJ...419..111E,1998ApJ...493..291B,2006A&A...450..887G}. Twisted magnetic flux tubes would form loops when they come into contact and magnetic field lines carried by colliding sections would be locally inverted. A current sheet will then form along the original contact plane and a pair of minijets will be perpendicular to the tube axes in the frame co-moving with the flux loop. The tube axes need to be very accurately aligned, otherwise the reconnection will proceed with a significant guide field, which, as we show later, is not a favourable situation.

\section{The structure of a single minijet}
\label{sec_struct}

\begin{figure}
  \includegraphics[width=\columnwidth]{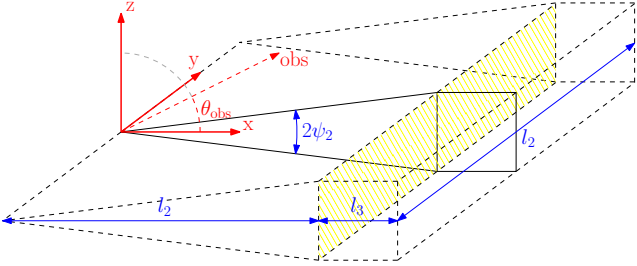}
  \caption{The geometric setup of our minijet model. The \emph{minijet region} (\emph{left}) is separated from the \emph{island region} (\emph{right}) by a stationary shock (\emph{yellow}). Reconnected plasma flows along $x$ axis. Spatial dimensions of the model are determined by the length $l_2$ and the opening angle $\psi_2$ of the minijet region.}
  \label{fig2}
\end{figure}

To describe the properties of a minijet, we adopt the scenario of relativistic Petschek reconnection \citep{1964NASSP..50..425P} with a guide field studied analytically by \cite{2005MNRAS.358..113L} and numerically by \cite{2009ApJ...705..907Z}. In the jet co-moving frame the current sheet is in the $xy$-plane, with the minijet outflow along $x$-axis (see Fig. \ref{fig2}). It is assumed that interaction between fast reconnected plasma and slow magnetic island leads to a stationary shock located at some $x=l_2$. This shock separates what we define as the \emph{minijet region} from the \emph{island region}. Parameters describing the jet flow are denoted with subscript `1`, those measured in the minijet region with subscript `2` and those measured in the island region with subscript `3`. The reference frame co-moving with the jet fluid is denoted by $\mathcal{O}_1$, the one co-moving with the minijet fluid by $\mathcal{O}_2$ and the one co-moving with the island fluid by $\mathcal{O}_3$. Quantities denoted by $''$ are measured in $\mathcal{O}_2$ and those with $'''$ in $\mathcal{O}_3$.

Jet plasma is assumed to be cold and highly magnetized ($\sigma_1\sim 100$). The magnetic field contains the antiparallel reconnecting component (within the $xz$-plane) of strength $B_1$ and the guide field (along the $y$-axis) of strength $B_{1,\rm G}=\alpha B_1$. According to the model of \cite{2005MNRAS.358..113L}, the reconnection outflow is thermal-pressure-dominated for $\alpha\lesssim 1/(2\sqrt{\sigma_1})$ and magnetically dominated otherwise. Although this limiting value of $\alpha$ is not dynamically important, it has a profound influence on the radiative properties of the system. In the following, we will consider models with no guide field (case I, $\alpha=0$) and models with a weak guide field (case II, $\alpha=1/(2\sqrt{\sigma_1})$). The inflow into the reconnection region is not relativistic, thus the relation between density and magnetic field is $\rho_1c^2\sim B_1^2/(4\pi\sigma_1)$.

\cite{2005MNRAS.358..113L} estimated the parameters of the reconnection outflow: bulk Lorentz factor $\Gamma_2\sim\sqrt{\sigma_1}$, density $\rho_2\sim 2\sqrt{\sigma_1}\rho_1$. The reconnected magnetic field is $B_2\sim B_1\theta_1/\sqrt{\sigma_1}$, where $\theta_1$ is the angle between the magnetic field lines and the oblique shock surface in the inflow region. However, if a guide field is present, it will be compressed to the value $B_{2,\rm G}\sim 2\sqrt{\sigma_1}B_{1,\rm G}$. In case II this yields $B_{2,\rm G}\sim B_1$, greatly exceeding the reconnected component. Pressure in the minijet region is given by $P_2=B_1^2/(8\pi)$ in case I. As demonstrated in numerical simulations by \cite{2009ApJ...705..907Z}, this parameter is very sensitive to the strength of the guide field. We adopt the following scaling: $P_2\sim (1-2\sigma_1\alpha^2)B_1^2/(8\pi)$, thus in case II the minijet pressure is roughly half of the value in case I. The ratio of thermal to rest energy densities is $P_2/(\rho_2c^2)\gtrsim\sqrt{\sigma_1}/8$, thus minijet matter is relativistically hot for $\sigma_1\gtrsim 60$. The magnetization of the minijet is $\sigma_2\sim\theta_1^2/(2\sigma_1)$ in case I and $\sigma_2\sim 1$ in case II. The minijet has an opening angle $\psi_2\sim \theta_1/(2\sigma_1)$ and volume $V_2\sim\psi_2l_2^3$, assuming that its width $\Delta y$ is similar to its length $l_2$.

The propagation of the minijet flow is affected by radiative processes described in Section \ref{sec_est_rad_min}: radiative losses of the electrons tend to reduce their pressure, while Compton drag by photons emitted from the island region causes deceleration of the bulk flow. Thus, we introduce correction factors that describe the final values of evolving parameters: $P_{\rm 2,f}=\eta_PP_2$ and $\Gamma_{\rm 2,f}=\eta_\Gamma\Gamma_2$. We will calculate these factors using the numerical scheme described in Section \ref{sec_num_sch}.

Initial parameters of the island region can be found by solving the shock-jump conditions. Under the assumption of negligible matter rest energy density on both sides of the shock and highly relativistic upstream flow, one can reduce the problem to a quadratic equation for $\Gamma_3$, giving the following solution (see Eq. 4.11 in \citealt{1984ApJ...283..694K}):
\bea
\Gamma_3^2 &\sim& \frac{1}{16(\sigma_{\rm 2,f}+1)}\left[8\sigma_{\rm 2,f}^2+26\sigma_{\rm 2,f}+17\right.+\nonumber\\
&+&\left.(2\sigma_{\rm 2,f}+1)\sqrt{16\sigma_{\rm 2,f}^2+16\sigma_{\rm 2,f}+1}\right]\,,
\eea
where $\sigma_{\rm 2,f} = B_2^2/(16\pi P_{\rm 2,f})$. Then we find:
\bea
\sigma_3 &\sim& \frac{\sigma_{\rm 2,f}}{(\sigma_{\rm 2,f}+1)\beta_3-\sigma_{\rm 2,f}}\,,\\
B_3 &\sim& \frac{\Gamma_{\rm 2,f}}{\Gamma_3\beta_3}B_2\,,\\
\rho_3 &\sim& \frac{\Gamma_{\rm 2,f}}{\Gamma_3\beta_3}\rho_2\,,\\
P_3 &\sim& \frac{B_3^2}{16\pi\sigma_3}\,.
\eea
Since $\Gamma_3\ll\Gamma_2$, radiation from the island region is strongly boosted in the minijet co-moving frame. And we expect that due to plasma compression both the magnetic field and average particle energy may be significantly higher than in the minijet region. We assume that the length of the island region is of the order of its height $l_3\sim 2\psi_2l_2$, hence its volume is $V_3\sim l_2l_3^2$.

\section{Emission model}
\label{sec_em}

\subsection{Particle energy distribution}

Non-thermal particle acceleration in relativistic plasmas is an open field of research. There are numerous studies of this process in relativistic shocks \citep[\eg,][]{1998PhRvL..80.3911B,2001MNRAS.328..393A,2003ApJ...589L..73L,2003ApJ...595..555N,2008ApJ...682L...5S,2009ApJ...698.1523S}, leading to a self-consistent (at least in weakly magnetized pair plasmas) scenario in which the first order Fermi process is initiated by the Weibel instability, producing power-law particle spectra $N\propto \gamma^{-p}$ of index $p\sim 2-3$. Particle acceleration in current sheets undergoing relativistic magnetic reconnection is in principle a more straightforward process, since strong electric fields are present. Numerical studies \citep{2003ApJ...586...72L,2004PhPl...11.1151J,2007ApJ...670..702Z,2008ApJ...677..530Z,2008ApJ...682.1436L} have shown that a relatively hard particle spectrum ($p\sim 1$) can be produced, if the current sheet can be stabilized against relativistic drift-kink instabilities (RDKI) by introduction of the guide component of magnetic field. Particles energized to the point that they are able to leave the current sheet can be further accelerated by the Fermi process, as they bounce between the two regions of reconnecting inflow \citep{2010arXiv1007.1522G}.

We are not investigating details of the particle acceleration process here, but consider the injection of relativistic electrons of fixed energy distribution. In the case of PKS 2155-304, the electron distribution cannot be constrained by multiwavelength observations if we accept the argument that X-ray emission is not produced co-spatially with the TeV emission. However, the \emph{H.E.S.S.} results imply the existence of a soft high-energy tail that in the flaring state shows a harder-when-brighter behaviour \citep{2010arXiv1005.3702H}.

The dynamical model of relativistic reconnection and the shock jump conditions provide us with the values of density $\rho$ and pressure $P$ in both the minijet and post-shock region. We assume that internal energy is equally divided between protons and electrons. The average energy of an electron depends on plasma composition and is highest when no electron-positron pairs are present. The electron number density is then given by $n_e\sim\rho/m_p$. The equation of state for both particle species is $P_{e(p)}=g_{e(p)}e_{e(p)}/3$, where $e$ is the internal energy density and $g$ is a parameter equal to 1 for relativistic particles and 2 for non-relativistic particles. From equipartition we have $e_e = e_p$, hence the pressure of electrons (protons) is
\be
P_{e(p)} \sim \frac{g_{e(p)}}{g_e+g_p}P\,,
\ee
while their average random Lorentz factor is
\be
\left<\gamma_{\rm e(p)}\right> \sim 1+\frac{e_{\rm e(p)}}{n_{\rm e(p)}m_{\rm e(p)}c^2} \sim 1+\frac{P}{\rho c^2}\frac{3}{g_e+g_p}\frac{m_p}{m_{\rm e(p)}}\,.
\ee
It is clear that electrons are highly and protons at least mildly relativistic, since $P/(\rho c^2)>1$. This points to a strongly peaked electron energy distribution.

For simplicity, we will adopt a relativistic Maxwellian electron distribution
\be
n_{\rm e}(\gamma) = n_{\rm e}\frac{\gamma^2}{2\gamma_0^3}e^{-\gamma/\gamma_0}
\ee
where $\gamma_0=\left<\gamma_{\rm e}\right>/3$. In models intended to fit the observed spectra, we add a power-law high-energy tail
\be
n_{\rm e}(\gamma) = n_{\rm e}\left\{
\begin{array}{l l}
\frac{\gamma^2}{2\gamma_0^3}e^{-\gamma/\gamma_0} & \gamma<(p+2)\gamma_0\\
\frac{\gamma^{-p}}{2\gamma_0^{1-p}}\left(\frac{p+2}{e}\right)^{p+2}e^{-\gamma/\gamma_{\rm cut}} & {\rm otherwise}
\end{array}
\right.\,,
\ee
where $p>2$ is the electron index,
\be
\gamma_0 = \left<\gamma_{\rm e}\right>\frac{2-\left[p^2+6p+10+\frac{(p+2)^3}{1-p}\right]e^{-(p+2)}}{6-\left[p^3+9p^2+30p+38+\frac{(p+2)^4}{2-p}\right]e^{-(p+2)}}
\ee
and $\gamma_{\rm cut}\gg (p+2)\gamma_0$ is the Lorentz factor of the exponential cut-off. For this choice of normalization factors and $\gamma_0$, the two distributions and their first derivatives are joined continuously. A different combination of relativistic Maxwellian and power-law components of the electron distribution has been proposed by \cite{2009MNRAS.400..330G}.

\subsection{Radiative processes}

In this Section, we calculate emission from the combined minijet and island regions. To this end, we consider synchrotron radiation, IC scattering (with Klein-Nishina cross-section) of local synchrotron photons and of radiation from the other emitting regions; radiative losses from all processes; Compton drag from anisotropic photon fields in the minijet co-moving frame and $\gamma\gamma$ pair production absorption. In Table \ref{tab_param} we compare parameter values in both cases I and II for reference values $\sigma_1=100$, $B_1=10\;{\rm G}$, $l_2=10^{14}\;{\rm cm}$ and $\Gamma\jet=10$ taken from \cite{2009MNRAS.395L..29G}.

\subsubsection{The minijet region}
\label{sec_est_rad_min}

Synchrotron emissivity is given by (see Eq. \ref{eq_em_syn_ave})
\be
j_{\rm SYN,2}'' = \frac{\sigma_Tc}{3\pi}n_{e,2}u_{\rm B,2}\left<\gamma_{e,2}^2\right>\,.
\ee
The photon energy of the synchrotron peak is at $h\nu_{\rm SYN,2}''\sim 2\times 10^{-8}\;(B_2/1\;{\rm G})\left<\gamma_{e,2}^2\right>\;{\rm eV}$. Energy density of the synchrotron radiation is given by (see Eq. \ref{eq_urad_patt2})
\be
\label{eq_u_syn2}
u_{\rm SYN,2}'' = \frac{j_{\rm SYN,2}''}{\Gamma_2c}\int_{V_2}\frac{dV}{r^2(1+\beta_2\cos\theta)}\,,
\ee
where $\theta$ is the angle to the emitting element measured from the flow velocity direction and $r$ is the distance to the emitting volume element. As the energy density of the synchrotron radiation is dominated by photons coming from the backwards direction, it should depend strongly on the position in the minijet region. We can estimate it, by taking into account only the solid angle of $\theta>(\pi-1/\Gamma)$, as $u_{\rm SYN,2}''\sim 2\psi_2j_{\rm SYN,2}''l_2/c$. Using this, we calculate synchrotron-self-Compton (SSC) emissivity $j_{\rm SSC,2}'' \sim j_{\rm SYN,2}''(u_{\rm SYN,2}''/u_{\rm B,2})$. The characteristic frequency of the SSC component is $h\nu_{\rm SSC,2}''\sim \left<\gamma_{e,2}^2\right>h\nu_{\rm SYN,2}''$. Klein-Nishina effects become important if $b_{\rm SSC,2}''=4\left<\gamma_{e,2}\right>h\nu_{\rm SYN,2}''/(m_ec^2)\gtrsim 1$. From Table \ref{tab_param}, we find that SSC emissivity is $\sim 3$ orders of magnitude weaker than synchrotron emissivity and is safely in the Thomson regime.

The energy density of radiation produced behind the stationary shock, in the island region, can be found from (see Eq. \ref{eq_urad_patt1}):
\be
\label{eq_u_ext2}
u_{3\to2}'' = \frac{\Gamma_2^2}{\Gamma_3^3}\frac{j_3'''}{c}\int_{V_3}\frac{(1+\beta_2\cos\theta)^2}{(1+\beta_3\cos\theta)^3}\frac{dV}{r^2}\,.
\ee
This radiation is observed mostly at small $\theta$, thus we can set $\cos\theta\sim 1$ to obtain an estimate (at $x=l_2/2$):
\be
u_{3\to2}'' \sim 16\pi\psi_2l_3\frac{\Gamma_2^2}{\Gamma_3^3(1+\beta_3)^3}\frac{j_3'''}{c}\,,
\ee
using which we find external Compton (EC) emissivity
\be
\frac{j_{\rm EC,2}''}{j_{\rm SYN,2}''} \sim \frac{3}{4}(1+\mu\obs'')^2\frac{u_{3\to2}''}{u_{\rm B,2}}\,,
\ee
where $\mu\obs''=(\cos\theta\obs+\beta_2)/(1+\beta_2\cos\theta\obs)$ is the cosine of the observer inclination in $\mathcal{O}_2$ and $\theta\obs$ is the observer inclination in $\mathcal{O}_1$ (see Section \ref{sec_obs_lum}). For the IC process only synchrotron photons from the island region are relevant; their characteristic energy in $\mathcal{O}_2$ is $h\nu_{3\to2}''\sim 2\Gamma_2/[\Gamma_3(1+\beta_3)]h\nu_{\rm SYN,3}'''$. The characteristic energy of EC photons is $\nu_{\rm EC,2}''\sim 2(1+\mu\obs'')\left<\gamma_{e,2}^2\right>\nu_{3\to2}''$ and the Klein-Nishina parameter $b_{\rm EC,2}''\sim h\nu_{\rm EC,2}''/(\left<\gamma_{e,2}\right>m_ec^2)$. We find that the EC component is stronger than the synchrotron component by $\sim 4$ orders of magniture in case I and 1 order of magnitude in case II. However, characteristic photon energies lie $\sim 2$ orders of magnitude within the Klein-Nishina regime and EC peaks should be suppressed. This means that in case II, synchrotron is likely to dominate.

From the relative importance of the spectral components we find that the radiative cooling of electrons is dominated by comptonization of the island photons in case I and by synchrotron emission in case II. Using Eq. (\ref{eq_cool}), we estimate the efficiency of electron cooling for plasma crossing the minijet length:
\be
\zeta_{\rm cool,2} = -\der{\gamma_e}{x}\frac{l_2}{\left<\gamma_{e,2}\right>} \sim \frac{4\sigma_T}{3m_ec^2}\frac{\left<\gamma_{e,2}\right>}{\Gamma_2}u_2''l_2\,,
\ee
where $u_2''$ stands for the largest of $u_{B,2}''$ and $u_{\rm EXT,2}''$. As shown in Table \ref{tab_param}, we obtain large cooling efficiency in case I, when $u_{\rm EXT,2}''\gg u_{\rm B,2}''$. However, as we are in the Klein-Nishina regime, this result is overestimated. In case II the cooling is not very efficient.

When electron cooling in the minijet region is dominated by IC losses off post-shock radiation, highly anisotropic in the minijet co-moving frame, and the minijet plasma is pressure-dominated, Compton drag may result in significant bulk braking of the minijet flow. The gradient of bulk Lorentz factor is given by (see Eq. 12 in \citealt{1996MNRAS.280..781S}):
\be
\label{eq_drag}
\left.\der{\Gamma_2}{x}\right|_{\rm drag} = -\frac{\int d\gamma\;n_{e,2}(\gamma)\;\dot{p}_{e,2}''(\gamma)}{u_{\rm e,2}+u_{\rm p,2}+u_{\rm B,2}}\,,
\ee
where $u_{\rm e(p),2}$ is the total co-moving energy density of electrons (protons), $u_{\rm B,2}=B_2^2/(8\pi)$ is the energy density of magnetic field and $\dot{p}_{e,2}''(\gamma)$ is the average co-moving radiative force per electron of random Lorentz factor $\gamma$. The full Klein-Nishina formula for $\dot{p}_e(\gamma)$ is provided in Appendix \ref{app_drag} (Eq. \ref{eq_drag_force_1e}). In the Thomson regime it becomes $\dot{p}_{e,2}''(\gamma)\sim(2/3)\sigma_T\gamma^2u_{\rm 3\to2}''$. We estimate the efficiency of the Compton drag as:
\be
\label{eq_zeta_drag}
\zeta_{\rm drag,2} = -\der{\Gamma_2}{x}\frac{l_2}{\Gamma_2} \sim \frac{2}{3}\frac{\left<\gamma_{e,2}^2\right>}{\Gamma_2}\frac{\left(n_{e,2}\sigma_Tl_2\right)u_{\rm 3\to2}''}{\rho_2c^2+3P_2+u_{\rm B,2}}\,.
\ee
We find that radiative drag should be more efficient in case I than in case II, however our estimate is again subject to Klein-Nishina suppression. We note that the SSC process can counter the radiation drag, as the local synchrotron radiation field is not isotropic in $\mathcal{O}_2$ and is actually stronger when arriving backwards with respect to fluid motion (see Eq. \ref{eq_u_syn2}). However, since external radiation dominates local synchrotron radiation, this effect is not important in our problem.  We left the detailed solution of this problem to our numerical scheme.

As a consequence of the deceleration of the minijet flow, plasma will be compressed in its co-moving frame. We adopt here adiabatic scaling of plasma parameters, noting that various plasma effects may complicate the relations among them. For a small decrease of Lorentz factor by $\delta\Gamma_2$, number density of particles should increase by $\delta n_2\sim (\delta\Gamma_2/\Gamma_2)n_2$. As the magnetic field is perpendicular, it increases like matter density $\delta B_2\sim(\delta\Gamma_2/\Gamma_2)B_2$. Pressure dominates the enthalpy and thus it should increase quadratically, $\delta P_2\sim 2(\delta\Gamma_2/\Gamma_2)P_2$. Such simultaneous changes in $n_2$ and $P_2$ imply an increase of energy of each electron by $\delta\gamma_{\rm e,2}\sim (\delta\Gamma_2/\Gamma_2)\gamma_{\rm e,2}$. Note that the electrons do not gain energy in the $\mathcal{O}_1$ frame, rather a part of their bulk kinetic energy is converted to random kinetic energy.

High-energy photons will be absorbed in photon-photon pair production mainly on the synchrotron radiation. The cross-section given in Eq. (\ref{eq_sigma_pp}) peaks at $\sigma_{\gamma\gamma}\sim \sigma_T/5$ for photon energy $h\nu_{\gamma\gamma,\rm peak}\sim 3.5(m_ec^2)^2/(h\nu_{\rm target})$. In $\mathcal{O}_1$, the minijet synchrotron photons provide targets of energy $h\nu_{\rm target,2}\sim 2\Gamma_2\,h\nu_{\rm SYN,2}''$ (as they come mainly from the backwards direction) and energy density $u_{\rm target,2}\sim 4\Gamma_2^2u_{\rm SYN,2}''$, while the island synchrotron photons have higher energy $h\nu_{\rm target,3}\sim h\nu_{\rm SYN,3}'''/[\Gamma_3(1+\beta_3)]$ and energy density $u_{\rm target,3}\sim u_{3\to2}''/(4\Gamma_2^2)$. Using Eq. (\ref{eq_kappa_pp}), we estimate the mean free path for high-energy photons
\be
\lambda_{\gamma\gamma,2(3)} = \frac{1}{\kappa_{\gamma\gamma,2(3)}} \sim \frac{5\,h\nu_{\rm target,2(3)}}{\sigma_T\,u_{\rm target,2(3)}}\,.
\ee
In Table \ref{tab_param} we find, that $\lambda_{\gamma\gamma,2(3)}\gg l_2$, thus we expect little absorption from both sources of synchrotron photons. However, when increasing the source characteristic size or parameters governing the synchrotron emissivity, we expect this effect may put important constraints on VHE luminosity. Detailed calculation will be performed in our numerical scheme.

\begin{table}
\centering
\caption{Parameters of plasma composition, electron distribution and radiative output from the minijet (subscript 2) and island (subscript 3) regions calculated for $\sigma_1=100$, $B_1=10\;{\rm G}$, $l_2=10^{14}\;{\rm cm}$, $\theta_1=0.5$, $\eta_P=1$, $\eta_\Gamma=1$, $\theta\obs=0.1$ and $\Gamma\jet=10$. Model with no guide field (case I) is compared to a model with a weak guide field (case II).}
\label{tab_param}
\begin{tabular}{c||c c}
case & I & II \\
\hline\hline
$\Gamma_2$ & \multicolumn{2}{c}{$10$} \\
$\Gamma_3$ & $1.06$ & $1.46$ \\
$\sigma_2$ & $1.25\times 10^{-3}$ & $1$ \\
$\sigma_3$ & $4\times 10^{-3}$ & $2.2$ \\
$B_2\;{\rm [G]}$ & $0.5$ & $10$ \\
$B_3\;{\rm [G]}$ & $14$ & $94$ \\
$\rho_2c^2\;{\rm [erg\,cm^{-3}]}$ & \multicolumn{2}{c}{$1.6$} \\
$\rho_3c^2\;{\rm [erg\,cm^{-3}]}$ & $45$ & $15$ \\
$P_2\;{\rm [erg\,cm^{-3}]}$ & $4$ & $2$ \\
$P_3\;{\rm [erg\,cm^{-3}]}$ & $1050$ & $80$ \\
\hline
$n_{e,2}\;{\rm [cm^{-3}]}$ & \multicolumn{2}{c}{$10^3$} \\
$n_{e,3}\;{\rm [cm^{-3}]}$ & $3\times 10^4$ & $10^4$ \\
$\left<\gamma_{e,2}\right>$ & $7000$ & $3500$ \\
$\left<\gamma_{e,3}\right>$ & $65000$ & $15000$ \\
$\zeta_{\rm cool,2}$ & $11.5$ & $0.8$ \\
$\zeta_{\rm cool,3}$ & $1.7$ & $1.9$ \\
$\zeta_{\rm drag,2}$ & $2.5$ & $0.1$ \\
\hline
$j_{\rm SYN,2}''\;{\rm [erg\,cm^{-3}\,s^{-1}]}$ & $10^{-6}$ & $10^{-4}$ \\
$j_{\rm SSC,2}''\;{\rm [erg\,cm^{-3}\,s^{-1}]}$ & $1.9\times 10^{-9}$ & $5\times 10^{-8}$ \\
$j_{\rm EC,2}''\;{\rm [erg\,cm^{-3}\,s^{-1}]}$ & $0.05$ & $1.7\times 10^{-3}$ \\
\hline
$h\nu_{\rm SYN,2}''\,{\rm [eV]}$ & $0.5$ & $2.4$ \\
$h\nu_{\rm SSC,2}''\,{\rm [eV]}$ & $2.3\times 10^7$ & $2.8\times 10^7$ \\
$h\nu_{\rm EC,2}''\,{\rm [eV]}$ & $3\times 10^{12}$ & $1.6\times 10^{11}$ \\
$b_{\rm SSC,2}''$ & $0.026$ & $0.06$ \\
$b_{\rm EC,2}''$ & $900$ & $90$ \\
\hline
$j_{\rm SYN,3}'''\;{\rm [erg\,cm^{-3}\,s^{-1}]}$ & $2.1$ & $1.6$ \\
$j_{\rm SSC,3}'''\;{\rm [erg\,cm^{-3}\,s^{-1}]}$ & $13.6$ & $0.13$ \\
$j_{\rm EC,3}'''\;{\rm [erg\,cm^{-3}\,s^{-1}]}$ & $3\times 10^{-8}$ & $1.9\times 10^{-9}$ \\
\hline
$h\nu_{\rm SYN,3}''\,{\rm [eV]}$ & $1200$ & $400$ \\
$h\nu_{\rm SSC,3}''\,{\rm [eV]}$ & $5\times 10^{12}$ & $9\times 10^{10}$ \\
$h\nu_{\rm EC,3}''\,{\rm [eV]}$ & $1.4\times 10^{8}$ & $7\times 10^{6}$ \\
$b_{\rm SSC,3}''$ & $600$ & $50$ \\
$b_{\rm EC,3}''$ & $4\times 10^{-3}$ & $9\times 10^{-4}$ \\
\hline
$h\nu_{\gamma\gamma,\rm peak,2}\,{\rm [eV]}$ & $10^{11}$ & $2\times 10^{10}$ \\
$h\nu_{\gamma\gamma,\rm peak,3}\,{\rm [eV]}$ & $1.1\times 10^9$ & $6\times 10^9$ \\
$\lambda_{\gamma\gamma,2}\,{\rm [cm]}$ & $1.6\times 10^{16}$ & $8\times 10^{14}$ \\
$\lambda_{\gamma\gamma,3}\,{\rm [cm]}$ & $2.6\times 10^{16}$ & $4\times 10^{16}$ \\
\hline
$L_{\rm LE,2}\;{\rm [erg\,s^{-1}]}$ & $5\times 10^{41}$ & $5\times 10^{43}$ \\
$L_{\rm LE,3}\;{\rm [erg\,s^{-1}]}$ & $2.7\times 10^{43}$ & $1.2\times 10^{44}$ \\
$L_{\rm HE,2}\;{\rm [erg\,s^{-1}]}$ & $2.3\times 10^{46}$ & $8\times 10^{44}$ \\
$L_{\rm HE,3}\;{\rm [erg\,s^{-1}]}$ & $1.8\times 10^{44}$ & $10^{43}$ \\
\end{tabular}
\end{table}

\subsubsection{The island region}
\label{sec_est_rad_isl}

Increased magnetic field and pressure will produce higher synchrotron emissivity than in the minijet region. Due to the assumed geometry of the region, synchrotron energy density can be estimated as $u_{\rm SYN,3}'''\sim\pi j_{\rm SYN,3}'''l_3/(2\Gamma_3c)$. Energy density of radiation from the minijet region is given by
\be
u_{2\to3}''' = \frac{\Gamma_3^2}{\Gamma_2^3}\frac{j_2''}{c}\int_{V_2}\frac{(1+\beta_3\cos\theta)^2}{(1+\beta_2\cos\theta)^3}\frac{dV}{r^2}\,.
\ee
In contrast to the situation in the minijet region, we have mainly $\theta\sim\pi$. Within the solid angle $\theta>\pi-1/\Gamma_2$, we can set $\cos\theta\sim -\beta_2$, obtaining an estimate
\be
u_{2\to3}''' \sim \frac{\pi\Gamma_2l_2}{2\Gamma_3^2(1+\beta_3)^2}\frac{j_2''}{c}\,.
\ee
The characteristic energy of photons from the minijet region is $h\nu_{2\to3}'''\sim 2\Gamma_2/[\Gamma_3(1+\beta_3)]\,h\nu_{\rm SYN,2}''$. EC emissivity differs by the sign of the cosine of scattering angle:
\be
\frac{j_{\rm EC,3}'''}{j_{\rm SYN,3}'''} \sim \frac{3}{4}(1-\mu\obs''')^2\frac{u_{2\to3}'''}{u_{\rm B,3}}\,,
\ee
where $\mu\obs'''=(\mu\obs+\beta_3)/(1+\beta_3\mu\obs)$. The characteristic energy of EC photons is $\nu_{\rm EC,3}'''\sim 2(1-\mu\obs''')\left<\gamma_{e,3}^2\right>\nu_{2\to3}'''$. Other parameters are found in the same way, as for the minijet region. We find that the strongest spectral components are synchrotron radiation and SSC, while the EC component is negligible. Synchrotron emissivity is similar in both cases, however SSC is much stronger in case I. The energies of SSC photons lie $\sim 2$ orders of magnitude into Klein-Nishina regime, similarily to EC photons in the minijet region.

The radiative cooling efficiency is calculated from
\be
\zeta_{\rm cool,3} = -\der{\gamma_e}{x}\frac{l_3}{\left<\gamma_{e,3}\right>} \sim \frac{4\sigma_T}{3m_ec^2}\frac{\left<\gamma_{e,3}\right>}{\Gamma_3}u_3'''l_3\,.
\ee
In case I cooling would be dominated by SSC ($\zeta_{\rm cool,3}\sim 1.7$), however Klein-Nishina suppression is likely to lead to the domination of synchrotron cooling ($\zeta_{\rm cool,SYN,3}\sim 0.26$). In case II synchrotron cooling is significant, so the electrons can radiate the bulk of their energy within the distance $l_3$.

\subsubsection{Observed luminosity}
\label{sec_obs_lum}

Let us place the observer in the $xz$ plane (see Fig. \ref{fig2}) at angle $\theta\obs$ to the minijet velocity direction as measured in jet rest-frame $\mathcal{O}_1$, so that $\bm{k}\obs=[\cos\theta\obs,0,\sin\theta\obs]$. The Doppler factor between the minijet co-moving frame $\mathcal{O}_2$ and $\mathcal{O}_1$ is $\mathcal{D}_{2\to1}=[\Gamma_2(1-\beta_2\cos\theta\obs)]^{-1}$, between the island co-moving frame $\mathcal{O}_3$ to $\mathcal{O}_1$ is $\mathcal{D}_{3\to1}=[\Gamma_3(1-\beta_3\cos\theta\obs)]^{-1}$ and between $\mathcal{O}_1$ and the laboratory frame is $\mathcal{D}_{1\to 0}=\Gamma\jet(1+\beta\jet\sin\theta\obs)$. The luminosity seen by the observer from the minijet (island) region is
\be
\label{L_obs}
L_{\rm obs,2(3)}=4\pi\mathcal{D}_{1\to0}^4\mathcal{D}_{2(3)\to1}^3j_{2}''(j_3''')V_{2(3)}\,.
\ee
In order for the island region to compete with the minijet region in observed luminosity, the emissivity ratio should be
\be
\frac{j_3'''}{j_2''} \sim \frac{\mathcal{D}_{2\to1}^3}{\mathcal{D}_{3\to1}^3}\frac{V_2}{V_3} \sim \left(\frac{\Gamma_2}{\Gamma_3}\right)^3\frac{1}{4\psi_2} \gtrsim 10^4 \,.
\ee
As can be seen in Table \ref{tab_param}, this is the case for the ratio of synchrotron emissivities, since $n_e$, $u_B$ and $\left<\gamma_e\right>$ are significantly higher in the island region. We find that the low-energy component from the island region should dominate in case I, while both components should be comparable in case II. However, high-energy emission is clearly dominated by Comptonization in the minijet region of synchrotron photons from the island region. Not only are the different spectral components produced by different mechanisms, but they also actually come from different (albeit adjacent) parts of the complex reconnecting system.

\subsubsection{Variability timescale}
\label{sec_var_tim}

The minijet length is constrained by the observed variability timescale. The $\gamma$-ray emission is dominated by the contribution from the minijet region, which is static in the jet co-moving frame $\mathcal{O}_1$. The variability timescale resulting from light-travel effects calculated in this frame should be divided by the Doppler factor of this frame with respect to observer, $\mathcal{D}_{1\to0}\sim \Gamma\jet$. The time delay between signals emitted by the same plasma portion at $x=0$ and $x=l_2$ is $\Delta t_x'=(l_2/c)(\beta_2^{-1}-\cos\theta\obs)$, which for $\theta\obs\sim\Gamma_2^{-1}$ is of the order of $\Delta t_x'\sim l_2/(\Gamma_2^2c)$. This would be larger if minijet plasma deceleration were taken into account. Time delays between signals emitted simultaneously in $\mathcal{O}_1$ across the minijet region width ($\Delta y\sim l_2$) are $\Delta t_y'=(l_2/c)\sin\theta\obs$; for $\theta\obs\sim\Gamma_2^{-1}$ it is $\Delta t_y'\sim l_2/(\Gamma_2c)$. Unless $\theta\obs\lesssim\Gamma_2^{-2}$, which is an unlikely situation, $\Delta t_y'\gg \Delta t_x'$ and the minijet length can be estimated as $l_2\lesssim \Gamma\jet\Gamma_2ct_{\rm var}$, which is consistent with the estimate for a spherical blob propagating freely in $\mathcal{O}_1$ with Lorentz factor $\Gamma_2$ \citep{2009MNRAS.395L..29G}.

\subsection{Numerical scheme}
\label{sec_num_sch}

We use exact formulae summarized in Appendix \ref{app_rad} to calculate the spectra of radiation emitted from the minijet and island regions. We use elements of the \emph{BLAZAR} code \citep{2003A&A...406..855M} with formulae valid in the Klein-Nishina regime \citep{2005MNRAS.363..954M}.

The minijet region is divided along the $x$ axis into $N$ sectors of width $(\Delta x)^i$, median position $x^i$ and volume $\Delta V^i=2\psi_2l_2\,(\Delta x)^i\,x^i$ fixed in $\mathcal{O}_1$. For each sector, the synchrotron emissivity spectrum $(j_{\rm SYN,2}'')^i(\nu'')$ is calculated from Eq. (\ref{eq_em_syn}), synchrotron energy density $(u_{\rm SYN,2}'')^i(\nu'')$ from Eq. (\ref{eq_u_syn2}), SSC emissivity $(j_{\rm SSC,2}'')^i(\nu'')$ from Eq. (\ref{eq_em_ic_iso_nu}), energy density of radiation from the island region
\be
(u_{3\to2}'')^i(\nu'') = \frac{\Gamma_2}{\Gamma_3^2}\frac{j_3'''(\nu''')}{c}\int_{V_3}\frac{1+\beta_2\cos\theta}{(1+\beta_3\cos\theta)^2}\frac{dV}{r^2}
\ee
(see Eq. \ref{eq_u_ext2}) and EC emissivity $(j_{\rm EC,2}'')^i(\nu'')$ from Eq. (\ref{eq_em_ic_mu_nu}). Energy densities of synchrotron radiation and the radiation from the island region are taken into account when calculating electron cooling rate, Compton drag and photon absorption.

Compton drag efficiency is calculated from Eqs. (\ref{eq_drag}) and (\ref{eq_zeta_drag}):
\be
\zeta_{\rm drag,2}^i = -\frac{(\Delta x)^i}{\Gamma_2^i}\times\left.\der{\Gamma_2^i}{x}\right|_{\rm drag}\,.
\ee
The evolution of minijet fluid parameters due to the plasma deceleration and compression is
\bea
\Gamma_2^i &\to& \left(1-\zeta_{\rm drag,2}^i\right)\Gamma_2^i\,,\\
B_2^i &\to& \left(1+\zeta_{\rm drag,2}^i\right)B_2^i\,,\\
n_2^i &\to& \left(1+\zeta_{\rm drag,2}^i\right)n_2^i\,,\\
P_2^i &\to& \left(1+2\zeta_{\rm drag,2}^i\right)P_2^i\,.
\eea

The electron distribution is initialized to $n_{e,2}$ in the first sector, then both effects of adiabatic compression and radiative cooling are taken into account in their evolution:
\bea
n_{e,2}^i(\gamma) &\to& \left(1+\zeta_{\rm drag,2}^i\right)n_{e,2}\left(\frac{\gamma}{1+\zeta_{\rm drag,2}^i}\right)\,,\\
n_{e,2}^i(\gamma) &\to& n_{e,2}^i(\gamma)+(\Delta x)^i\times\der{n_{e,2}^i(\gamma)}{x}\,,
\eea
where we employ a kinetic equation
\be
\der{n_{e,2}(\gamma)}{x} = -\pder{}{\gamma}\left(n_{e,2}(\gamma)\times\left.\der{\gamma}{x}\right|_{\rm rad}\right)
\ee
and $\der{\gamma}{x}|_{\rm rad}$ is the cooling rate given by Eq. (\ref{eq_cool}).

The evolution is tempered by including the effect of mixing the evolved fluid with freshly reconnected plasma proportionally to their volumes:
\be
X_2^{i+1} = \frac{\left[x^i-(\Delta x)^i/2\right]\times X_2^i+(\Delta x)^i\times X_2}{x^i+(\Delta x)^i/2}\,,
\ee
where $X_2$ stands for $\Gamma_2$, $B_2$, $n_2$, $P_2$ or $n_{\rm e,2}(\gamma)$.

Essentially, at every step we mix $x^i-(\Delta x)^i/2$ parts of \emph{old} electrons evolved via radiative losses with $(\Delta x)^i$ parts of \emph{new} electrons injected through the minijet boundary. The numerical method for solving the kinetic equation is explained in \cite{2003A&A...406..855M}.

From the final electron distribution $n_{e,2}^{N+1}$ we find $\left<\gamma_{e,2}\right>^{N+1}$, and then the pressure
\be
P_2^{N+1} = \left(\frac{\left<\gamma_{e,2}\right>^{N+1}}{\left<\gamma_{e,2}\right>}g_{e,2}+g_{p,2}\right)\frac{P_2}{g_{e,2}+g_{p,2}}\,.
\ee
Our approach to properly accounting for the effects of electron cooling and Compton drag is to find such values of corrections $\eta_P$ and $\eta_\Gamma$, for which $P_2^{N+1}=P_{\rm 2,f}$ and $\Gamma_2^{N+1}=\Gamma_{\rm 2,f}$.

In a similar way we calculate emissivities in the island region, which is also divided into several sectors, however the effects of Compton drag and plasma mixing are not considered in this case. We then calculate the luminosities from each sector and transform them to the laboratory frame, using Eq. (\ref{L_obs}).

Pair production opacity is calculated in $\mathcal{O}_1$ by following a ray emitted at the center of each sector in several steps evenly spaced over distance $2l_2$. The minijet region is divided into a 2-dimensional array $(j,k)\in(1,...,N)\times(1,...,N)$ of sectors centered at $x^j=(j-1/2)l_2/N$ and $y^k=[(k-1/2)/N-1/2]l_2$ of volume $\Delta V_2^{j,k}=2\psi_2l_2^2x^j/N^2$, while the island region is divided into a 1-dimensional array $(k)\in(1,...,N)$ of sectors centered at $x=l_2+l_3/2$ and $y^k=[(k-1/2)/N-1/2]l_2$ of volume $\Delta V_3^k=2\psi_2l_2^2l_3/N$. For each volume element, the mean distance $r_{\rm mean}$ from the photon, mean scattering angle and mean Doppler factor of the emitting fluid are calculated and used to determine energy density of incident radiation at given position. A problem arises when $r_{\rm mean}$ is small or comparable to the size $R$ of the volume element. For a spherical volume element containing static, isotropic and optically thin emitting fluid, the radiation density at distance $r_{\rm mean}$ from the center is given by
\be
u(r_{\rm mean})=\frac{3L}{8\pi R^2c}\left(1-\frac{q^2-1}{2q}\ln\left|\frac{q+1}{q-1}\right|\right)\,,
\ee
where $q=r_{\rm mean}/R$ and $L=4\pi jV(R)$ is the total luminosity. This formula can be well approximated for both small and large values of $q$ with the following:
\be
u(r_{\rm mean})\sim \frac{L}{4\pi(r_{\rm mean}^2+R^2/3)c}\,.
\ee
We thus replace the mean distance with $r^2=r_{\rm mean}^2+R^2/3$ and use formulae valid in the limit of large $q$. We adopt $R\sim l_2/(2N)$ for elements of both the minijet and island regions.

\section{Results}
\label{sec_res}

We begin by running our numerical scheme with the same parameters as used in Table \ref{tab_param}. We denote this model as Model A. Since effects like opacity, Compton drag and radiative cooling are weak, a direct comparison of emissivities, energy densities and luminosity components between analytical estimates and numerical results can be performed.

\begin{figure}
  \includegraphics[width=\columnwidth]{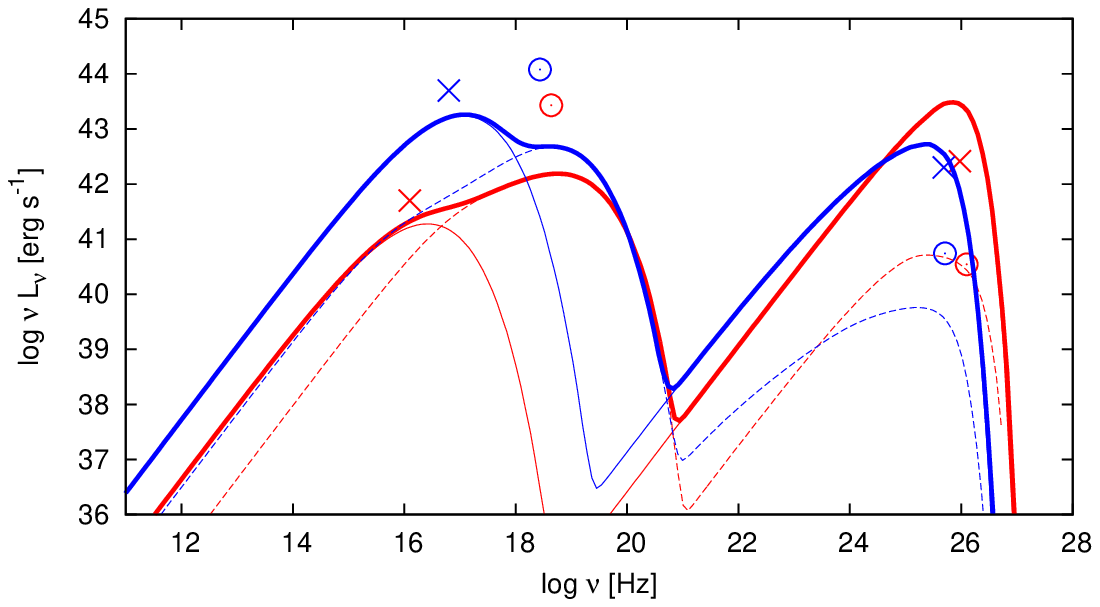}
  \includegraphics[width=\columnwidth]{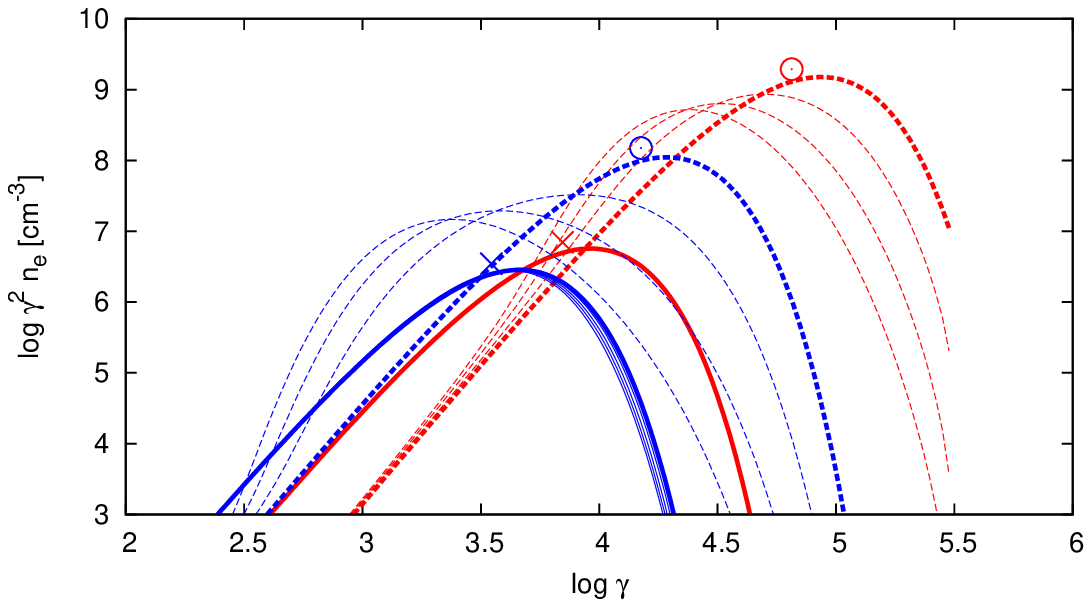}
  \includegraphics[width=\columnwidth]{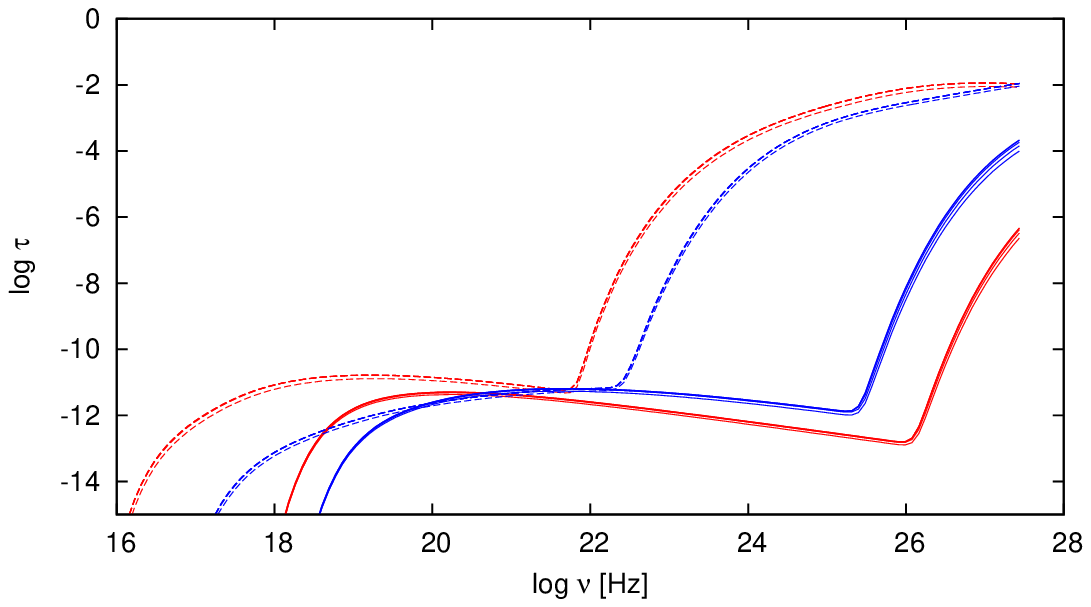}
  \caption{Results obtained for Model A -- parameter values are $\sigma_1=100$, $l_2=10^{14}\;{\rm cm}$ and $B_1=10\;{\rm G}$. The observer is located at $\theta\obs=0.1$. \emph{Red lines} denote model with no guide field (case I), while \emph{blue lines} denote model with weak guide field (case II). Analytical estimates from Table \ref{tab_param} are marked with \emph{crosses} (for the minijet region) and \emph{circles} (for the island region). {\bf Top panel}: SEDs of the minijet (\emph{thin solid lines}) and the island (\emph{thin dashed lines}) regions, as well as their sums (\emph{thick solid lines}), in the laboratory frame. {\bf Middle panel}: electron number density energy distributions in local co-moving frames. Evolution of the electron distribution is illustrated by plotting results for several sectors out of 10 equally spaced along $x$-axis. For the minijet region (\emph{solid lines}) we show sectors 2 (\emph{thick line}), 4, 6, 8 and 10 (lines are barely distinguishable as a result of inefficient cooling). For the island region (\emph{dashed lines}), we show sectors 1 (\emph{thick lines}), 2, 3 and 4. {\bf Bottom panel}: pair-production opacity for high-energy photons emitted from the minijet region contributed by soft photons from the minijet (\emph{solid lines}) and island (\emph{dashed lines}) regions. We plot opacities for photons emitted from sectors 2, 4, 6, 8 and 10 of the minijet region. The results for different sectors are barely distinguishable, thus we use the same line types.}
  \label{fig3}
\end{figure}

\begin{figure*}
  \includegraphics[width=0.49\textwidth]{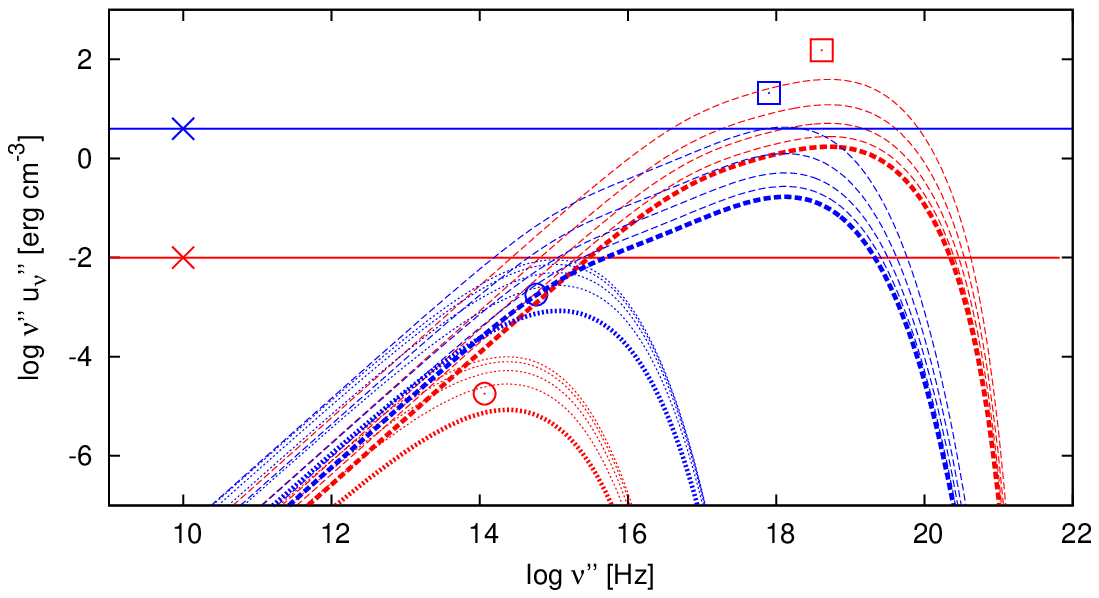}
  \includegraphics[width=0.49\textwidth]{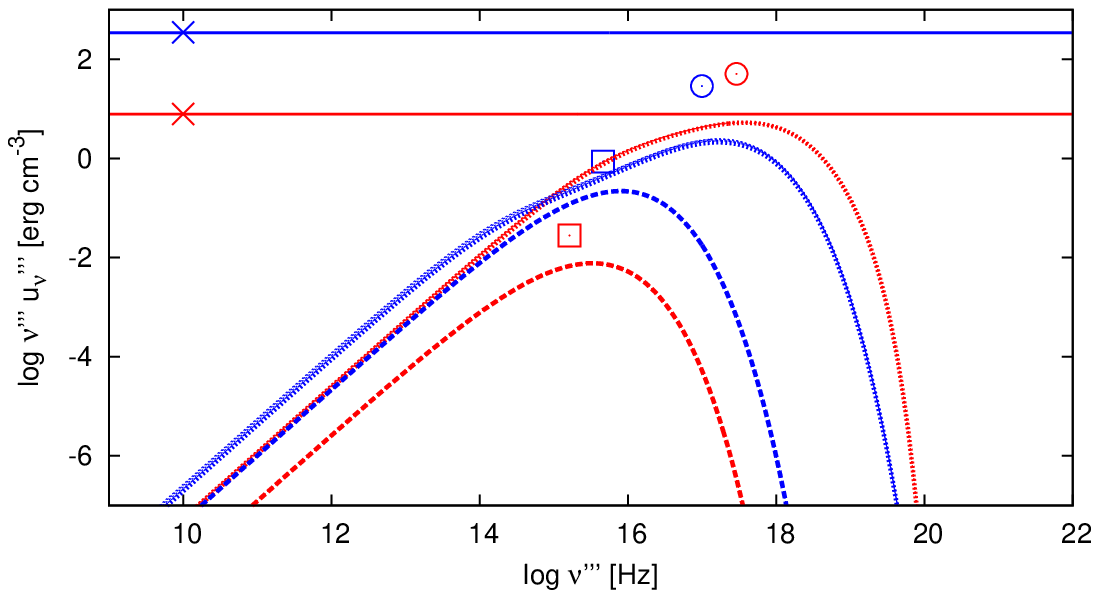}
  \includegraphics[width=0.49\textwidth]{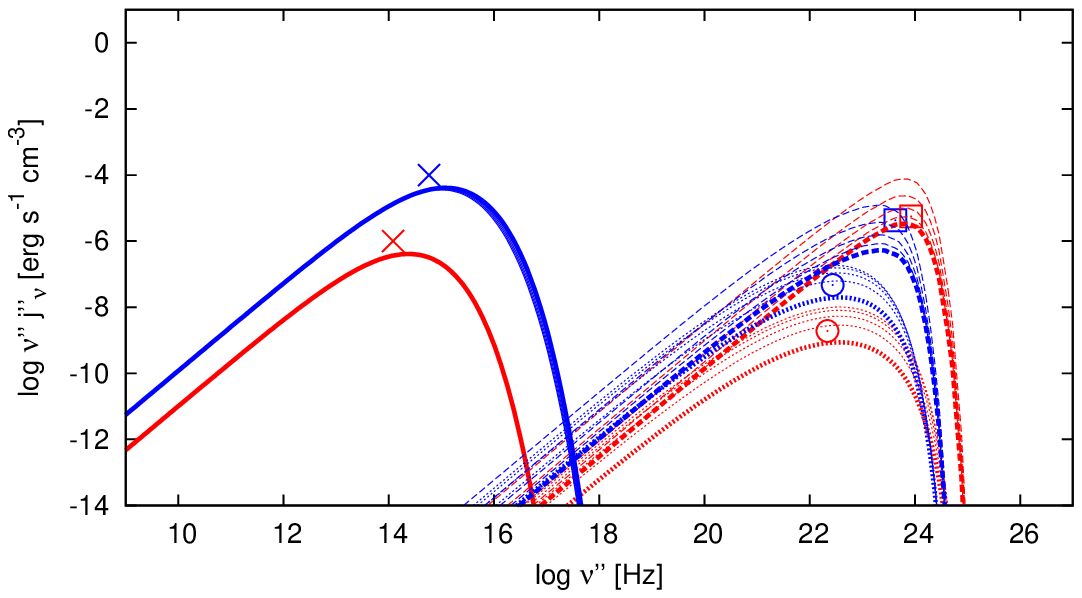}
  \includegraphics[width=0.49\textwidth]{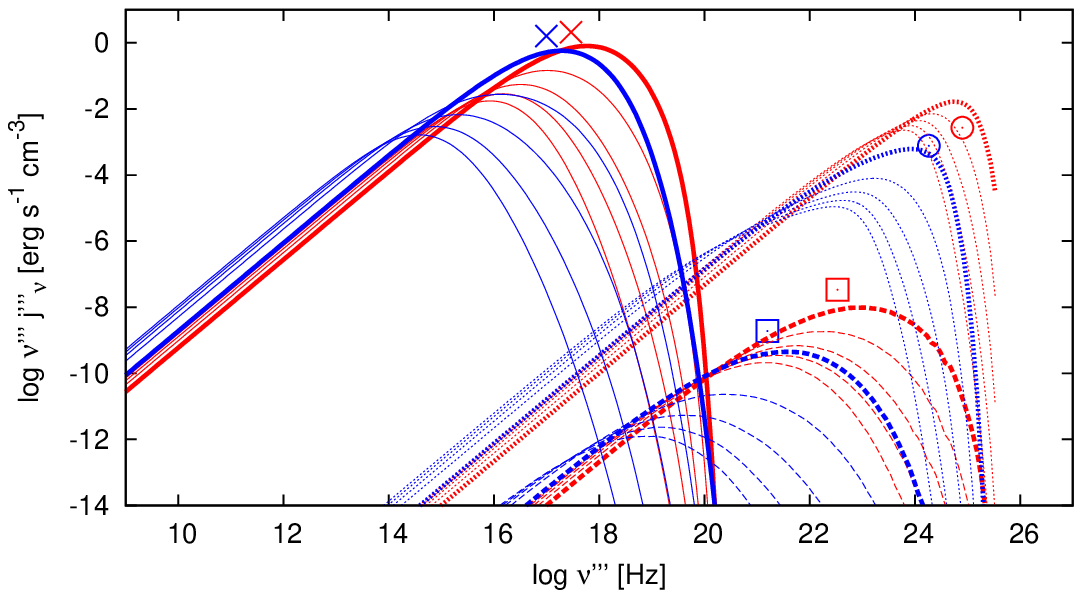}
  \caption{Local energy densities and emissivities for model A -- same as in Fig. \ref{fig3}. Analytical estimates from Table \ref{tab_param} are marked with symbols explained below. {\bf Left panels}: the minijet region -- sectors 2 (\emph{thick lines}), 4, 6, 8 and 10 out of 10. {\bf Right panels}: the island region -- sectors 1 (\emph{thick lines}), 3, 5, 7 and 9 out of 10. {\bf Top panels}: energy density of the magnetic field (\emph{solid lines} and \emph{crosses}), synchrotron radiation (\emph{dotted lines} and \emph{circles}) and external radiation (\emph{dashed lines} and \emph{squares}). {\bf Bottom panels}: synchrotron (\emph{solid lines} and \emph{crosses}), SSC (\emph{dotted lines} and \emph{circles}) and EC (\emph{dashed lines} and \emph{squares}) emissivities, all calculated in their respective local frames. \emph{Line colours} have the same meaning as in Fig. \ref{fig3}.}
  \label{fig4}
\end{figure*}

In Fig. \ref{fig3} we present spectral energy distributions (SEDs) in the observer's frame, electron energy distributions and pair production opacity as a function of emitted frequency. There is good agreement between estimated and numerically calculated positions and values of peak radiation energy densities, local emissivities and observed luminosities. Synchrotron emission from the island region has been overestimated, because effective cooling of electrons reduces the average emissivity. The EC component from the minijet region has been underestimated, because emissivity from the final minijet sector rapidly increases in the vicinity of the stationary shock front. Cooling efficiency is low for the minijet electrons and high for the island electrons. The pair production opacity for TeV photons is at a comparable level of $\tau\sim 0.01$ in both cases and is very uniform when measured in different geometrical sectors. It is dominated by the contribution from the island synchrotron photons, since absorption by minijet synchrotron photons targets higher-energy emission. This is in qualitative agreement with target energies and mean-free paths listed in Table \ref{tab_param}.

In Fig. \ref{fig4} we show radiation energy densities and emissivities in local frames. There is strong evolution of the synchrotron emissivity from the island region (solid lines in bottom right panel), reflected in the local synchrotron radiation energy density in the island region (dotted lines in top right panel) and the external radiation energy density in the minijet region (dashed lines in top left panel). This, however, does not affect the dominant high-energy emissivity components: EC in the minijet region (dashed lines in bottom left panel) and SSC in the island region (dotted lines in bottom right panel), which agree with the estimates.

\begin{figure}
  \includegraphics[width=\columnwidth]{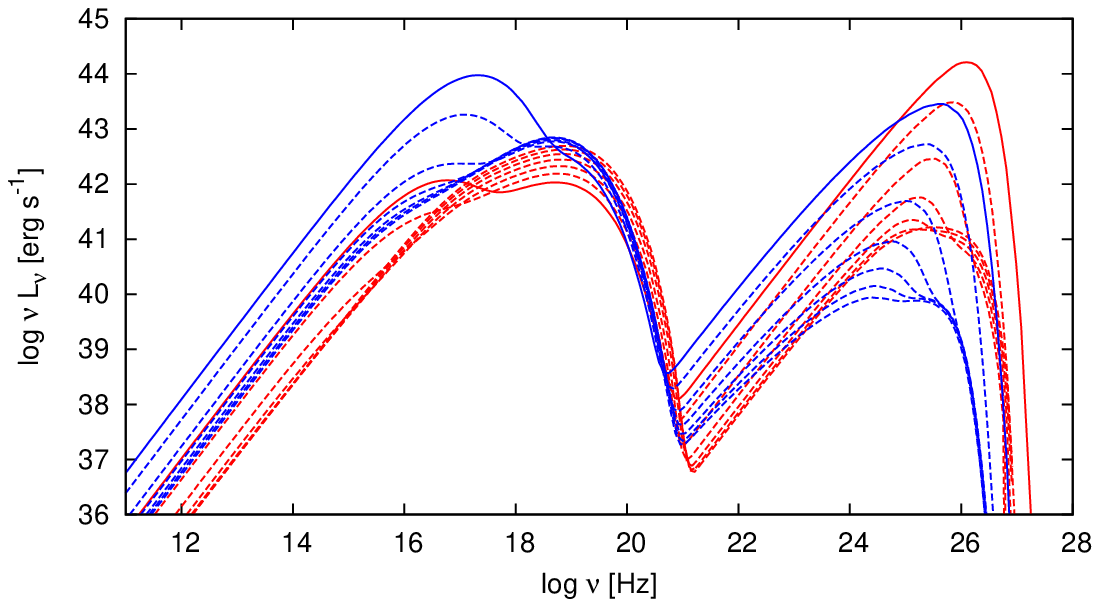}
  \includegraphics[width=\columnwidth]{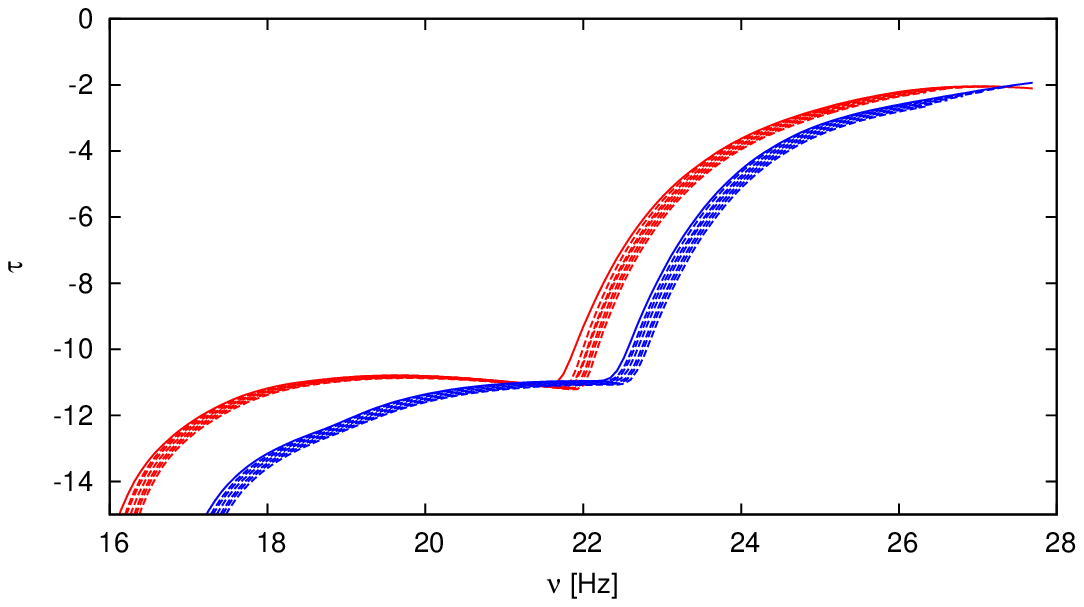}
  \caption{The effect of changing the observer's inclination angle $\theta\obs$, as measured with respect to the flow velocity ($x$-axis) in the jet co-moving frame $\mathcal{O}_1$, on the observed SED ({\bf top panel}; summed emission from the minijet and island regions) and pair production opacity ({\bf bottom panel}; for the minijet photons emitted from sector 10; by all soft photons) for Model A. The values of $\theta\obs$ are 0 (\emph{solid lines}), 0.1 (default value for other models), ..., 0.6 (\emph{dashed lines}). \emph{Line colours} have the same meaning as in Fig. \ref{fig3}.}
  \label{fig5}
\end{figure}

In Fig. \ref{fig5} we show the effect of varying the observer's inclination angle $\theta\obs$ on observed SED and opacity. For the values of this angle of the order of a few times $1/\Gamma_2 \sim 0.1$, the Doppler factor of the minijet region $\mathcal{D}_{2\to1}$ varies much more than the Doppler factor of the island region $\mathcal{D}_{3\to1}$. The flux produced in the minijet region, which dominates in soft X-ray and GeV-TeV bands, decreases strongly with increasing $\theta\obs$. On the contrary, the hard X-ray flux originating in the island region slightly increases, because the Doppler factor of the jet $\mathcal{D}_{1\to0}$ increases faster than $\mathcal{D}_{3\to1}$ decreases. For $\theta\obs\gtrsim 0.5$, emission from the island region dominates the total output. Pair-production opacity shows variation of threshold energy with increasing $\theta\obs$: a moderate decrease for the contribution from the minijet photons and a small increase for the contribution from the island photons. However, the contribution from the island photons always dominates and the net effect for TeV photons is a small decrease of opacity with increasing $\theta\obs$.

\subsection{Maximum TeV luminosity}

\cite{2009MNRAS.395L..29G} chose the parameter values for Model A in order to obtain a TeV luminosity of $\sim 10^{47}\;{\rm erg\;s^{-1}}$, as required for observed fast TeV flares. The emitting region size $l_2\sim 10^{14}\;{\rm cm}$ was estimated from energetic considerations corresponding to a jet magnetic field value of $B_1\sim 12\;{\rm G}$, while the variability timescale constraint was $l_2\lesssim 9\times 10^{14}\;{\rm cm}$ for $t_{\rm var}=5\;{\rm min}$. We find that $\gamma$-ray luminosities for Model A are of the order of $10^{43}\;{\rm erg\;s^{-1}}$, $\sim 4$ orders of magnitude lower than required. This discrepancy likely comes from the different geometrical shape of the emitting region. \cite{2009MNRAS.395L..29G} used a highly idealized blob model for the emitting region, but a more realistic description of relativistic magnetic reconnection requires extremely small opening angles of the emitting regions, which limits the emitting volume. To increase the luminosity, we could increase either the energy density regulated by $B_1$ and/or the minijet length. We will therefore consider models with a minijet length approaching the maximum allowed by the variability timescale and tune the value of the jet magnetic field to produce the desired value of the luminosity.

It is then interesting to determine the maximum TeV luminosity that can be produced by minijets and to identify the physical effect that limits it. An obvious limitation on $\gamma$-ray luminosity comes from the intrinsic opacity to high-energy radiation. Another constraint comes from radiative efficiency. If the electrons are not heated everywhere accross the source, the effective emitting volume will depend on the average cooling distance of high-energy electrons. In our model, efficient Compton drag could also be a factor limiting high-energy emission, as it decreases the boosting factor for part of the emitting volume. All these effects become more prominent with increasing energy density of soft radiation and our numerical scheme is designed to deal with such a problem.

\begin{figure}
  \includegraphics[width=\columnwidth]{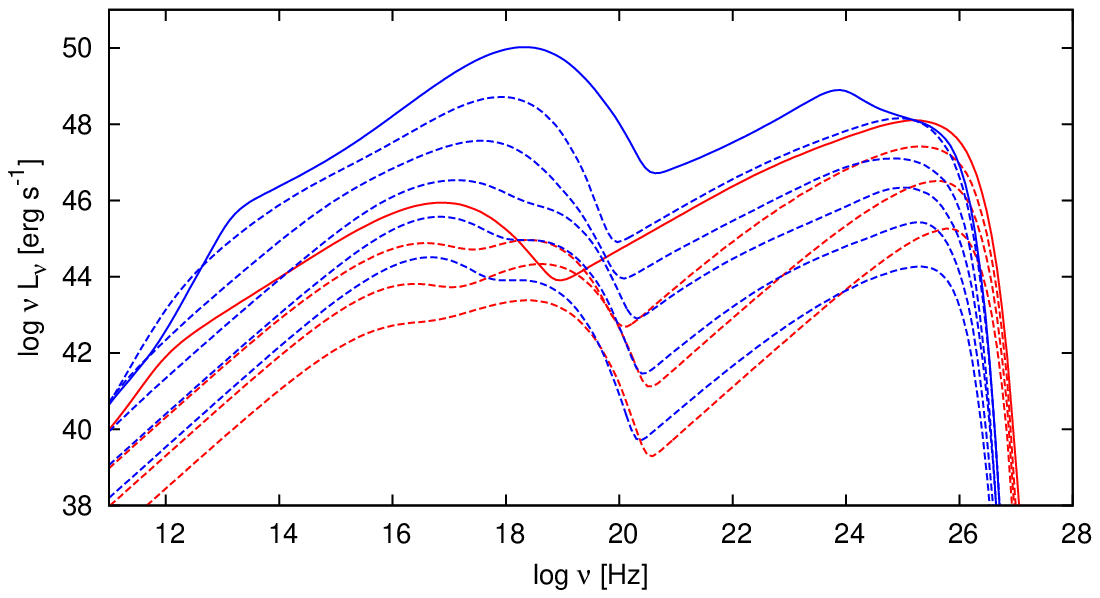}
  \includegraphics[width=\columnwidth]{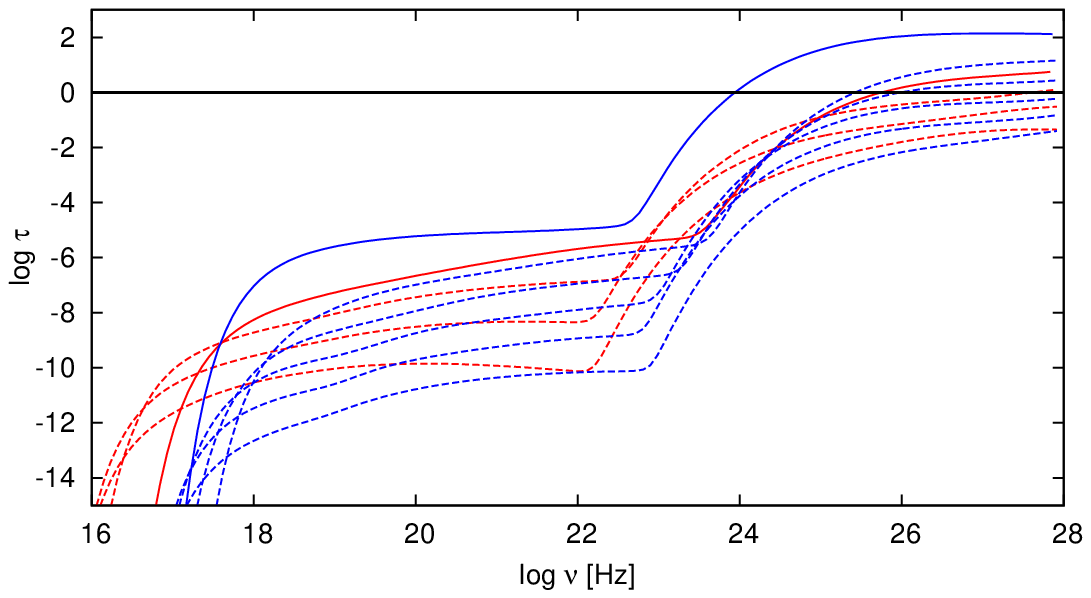}
  \includegraphics[width=\columnwidth]{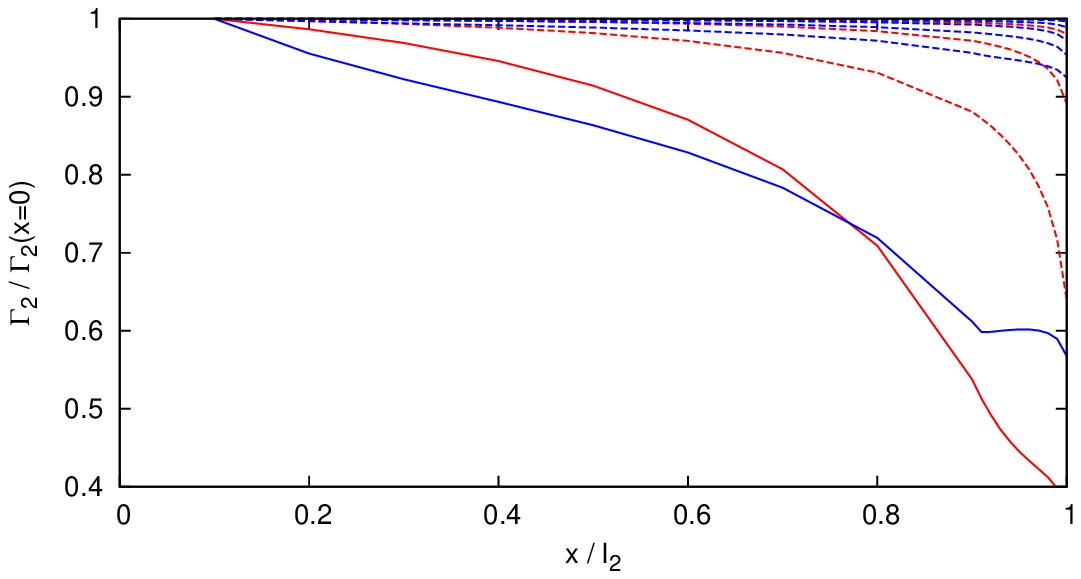}
  \caption{The effect of increasing the minijet energy density, related by the reconnection model to the jet magnetic field strength $B_1$, on the observed SED ({\bf top panel}; summed emission from the minijet and island regions), pair production opacity ({\bf middle panel}; all contributions summed for the minijet photons emitted at $x=0.9 l_2$) and minijet dynamics ({\bf bottom panel}; evolution of the Lorentz factor along the minijet region). The models have been calculated for minijet length $l_2=9\times 10^{14}\;{\rm cm}$. The values of $B_1$ (in Gauss) are: in case I (\emph{red lines}) -- 4, 8, 16 (\emph{dashed lines}) and 32 (\emph{solid line}); in case II (\emph{blue lines}) -- 4, 8, 16, 32, 64 (\emph{dashed lines}) and 128 (\emph{solid line}).}
  \label{fig6}
\end{figure}

The relativistic Petschek reconnection model from \cite{2005MNRAS.358..113L} relates particle density and pressure to the magnetic field strength, thus we can increase radiative energy densities and emissivities by changing one parameter, without altering the geometric structure. In Fig. \ref{fig6}, we show a series of models with increasing jet magnetic field. We find that on average, an increase of magnetic field by a factor $2$ increases both synchrotron and IC luminosities by a factor $\sim 11$ and opacity at $\nu=10^{26}\;{\rm Hz}$ by a factor $\sim 5$. The opacity approaches unity when high-energy luminosity is at the level of $10^{47-48}\;{\rm erg\;s^{-1}}$. Cases I and II do not differ significantly in terms of luminosity (with the exception of the soft X-ray synchrotron peak from the minijet region) and opacity, however, Compton drag is more efficient in case I, as already predicted in Section \ref{sec_est_rad_min}. For the models with opacity close to unity, the bulk Lorentz factor in the minijet region drops by $\sim 60\%$ in case I and only by $\sim 5\%$ in case II. Thus, the Compton drag effect can be an important factor limiting high-energy luminosity in Compton-dominated anisotropic sources.

The conclusion from the previous paragraph is that for a minijet size of $10^{14}\;{\rm cm}$, jet magnetization $\sigma_1=100$ and bulk Lorentz factor $\Gamma\jet=10$, the $\gamma$-ray luminosity is limited by opacity at $\sim 10^{48}\;{\rm erg\;s^{-1}}$. Can we relax this constraint by considering the Comptonization of external radiation? Note that our scenario suggests a chain of minijets forming a ring-like structure. Each minijet should have two neighbours, one sharing the X-point and one sharing the O-point. Both would be directed away from the external observer and thus relativistically hidden. The average inclination angle between two neighbours is inversely proportional to the number $N_{\rm ring}$ of individual minijets in the ring. If $N_{\rm ring}\gtrsim \pi\Gamma_2$, radiation from the O-point-sharing neighbour is strongly boosted in the comoving frame of the observed minijet region ($\mathcal{O}_2$). It is almost completely anisotropic and can dominate the Compton drag effect.

\begin{figure}
  \includegraphics[width=\columnwidth]{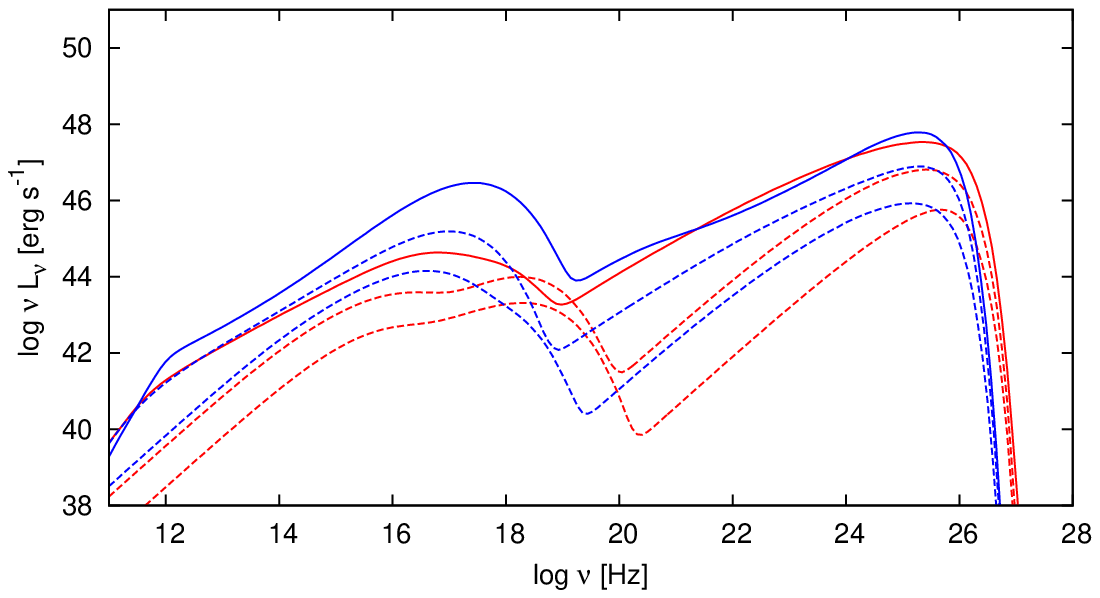}
  \includegraphics[width=\columnwidth]{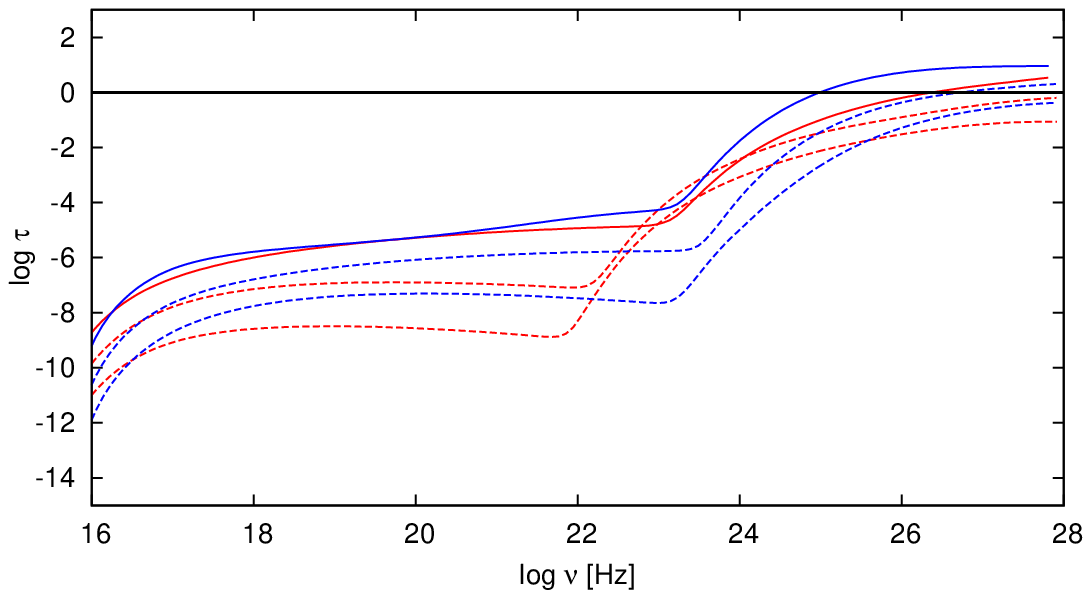}
  \includegraphics[width=\columnwidth]{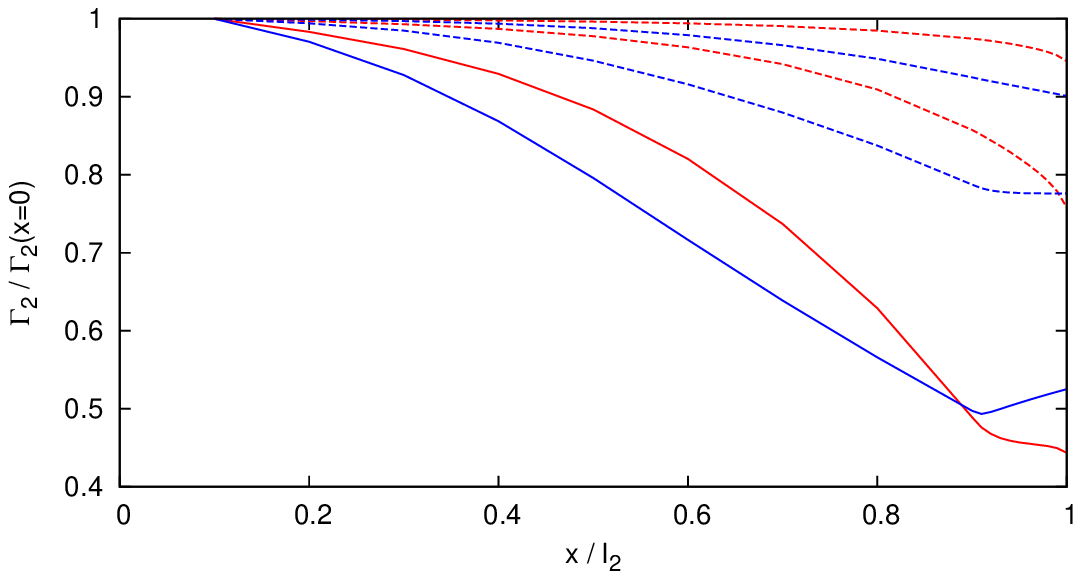}
  \caption{Same as in Fig. \ref{fig6}, but taking into account radiation from an opposite minijet system. The values of $B_1$ (in Gauss) in both cases are: 4, 8 (\emph{dashed lines}) and 16 (\emph{solid line}).}
  \label{fig7}
\end{figure}

In Fig. \ref{fig7} we show a series of models taking into account the radiative interaction with an opposite minijet system. It is assumed that the opposite minijet has identical parameters and is also affected by interaction with observed minijet. We find that the major difference that this additional radiation makes is to increase the efficiency of Compton drag. In the case of a single minijet we only calculated the drag effect from synchrotron photons originating in the island region. They had relatively high energy and the bulk of them were scattered in the Klein-Nishina regime. Radiation from an oppositely-directed minijet is highly anisotropic, regardless of the region from which it was emitted. Photons from the minijet region have lower energies and are more effectively scattered, even though they are especially strongly boosted in $\mathcal{O}_2$ frame.

Strong Compton drag limits the maximum magnetic field strength for which our model is valid, {\ie}, for which the deceleration is not catastrophic. This limits the resulting X-ray luminosity (which is of synchrotron origin), especially in case II. Because of higher synchrotron emission from the minijet region, models with a guide field are more sensitive to the existence of an opposite minijet system. Comparing results for models calculated with $B_1=4\;{\rm G}$, the differences between isolated and mirrored minijets are small in case I, while in case II we observe a stronger IC component, higher opacity and noticeable drag when radiation from an oppositely-directed minijet is present.

For the brightest models that we have obtained, the $\gamma$-ray luminosity is of the same order of magnitude as in the case of an isolated minijet, $\sim 10^{48}\;{\rm erg\;s^{-1}}$. For models with a guide field (case II), this can be achieved with much lower synchrotron emission, {\ie}, larger Compton dominance. This is an important advantage in light of observational properties of the best-studied fast TeV outburst.

\subsection{Application to TeV blazars}

We apply our model to a TeV flare of PKS 2155-304 observed by \emph{H.E.S.S.} in July 2006, which reached luminosities $~10^{46}\;{\rm erg\,s^{-1}}$ \citep{2007ApJ...664L..71A}. We are additionally constrained by the simultaneous \emph{Chandra} observations in the soft X-ray band \citep{2009A&A...502..749A}. Since the variability amplitude is much lower in the X-ray band and there is evidence for an almost cubic relation between TeV and X-ray fluxes, the X-ray luminosity of the minijet must stay below the \emph{Chandra} result.

We are attempting to match the observed TeV spectrum of photon index $\sim 3.5$ (above $\sim 0.4\;{\rm TeV}$), thus a power-law tail is required in the electron distribution of the minijet region. Because this part of the IC spectrum is produced in the Klein-Nishina regime, there is no straightforward relation between the electron spectral index and the photon index. We find that the value $p=3.2$ ($N_\gamma\propto \gamma^{-p}$) produces an adequate slope of the TeV spectrum.

\begin{figure}
  \includegraphics[width=\columnwidth]{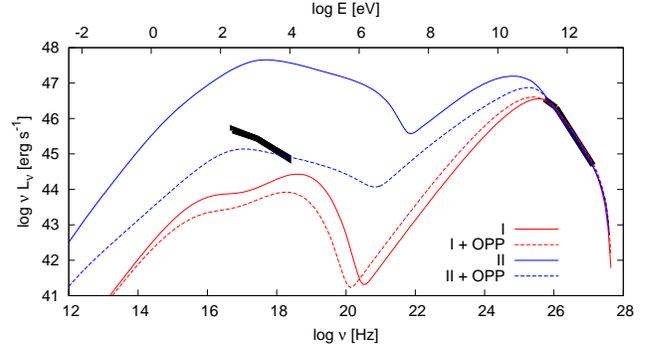}
  \caption{SEDs of the minijet models matching the TeV spectrum of flaring state of PKS 2155-304, compared with July 2006 simultaneous observations by \emph{H.E.S.S.} and \emph{Chandra} (\emph{thick black lines}). \emph{Red lines} show models with no guide field (case I), \emph{blue lines} models with significant guide field (case II), \emph{solid lines} models of isolated minijets and \emph{dashed lines} models with radiation from an opposite minijet system (`OPP`).}
  \label{fig8}
\end{figure}

In Fig. \ref{fig8}, we show four models with TeV luminosity matching the \emph{H.E.S.S.} observation of PKS 2155-304: with (`II`) or without (`I`) a guide field and with (`+OPP`) or without opposite minijet radiation. All models have been calculated for minijet region size $l_2=9\times 10^{14}\;{\rm cm}$. The jet magnetic field values $B_1$ for this models are 8.5 G (`I`), 35 G (`II`), 7 G (`I+OPP`) and 8 G (`II+OPP`). High-energy components are very similar for all models, while low-energy components are widely diverse with peak luminosities spanning 4 orders of magnitude. More soft radiation is produced in models with a guide field and without opposite minijet radiation. The low-energy component is broader in the presence of a guide field, extending up to $\sim 100\;{\rm MeV}$ instead of $\sim 1\;{\rm MeV}$ without a guide field. The guide field affects also the spectral shape of the low-energy component in the hard X-ray band: it is sharply peaked in case I, while in case II it shows a soft power-law plateau. The constraints imposed by \emph{Chandra} observations exclude model `II`, which severely overproduces X-ray flux. Model `II+OPP` is marginally consistent with the data at the high-energy end of the observational band.

As already mentioned in the previous subsection, a single current sheet may produce more than one minijet at the same time. Since there is no preference for any direction in the plane perpendicular to the jet (which is true for a ring-like structure), individual minijets would arise with different orientations and most of them would be strongly misaligned with respect to the line of sight. As we showed in Fig. \ref{fig5}, radiation from the minijet is strongly anisotropic, but relativistic boosting is much stronger in the minijet region, which dominates $\gamma$-ray emission, than in the island region, which produces strong hard X-ray emission. Emission from misaligned minijets should be taken into account in a discussion of X-ray constraints. They can also contribute to the $\gamma$-ray luminosity when $N_{\rm ring}\gg\pi\Gamma_2$.

\begin{figure}
  \includegraphics[width=\columnwidth]{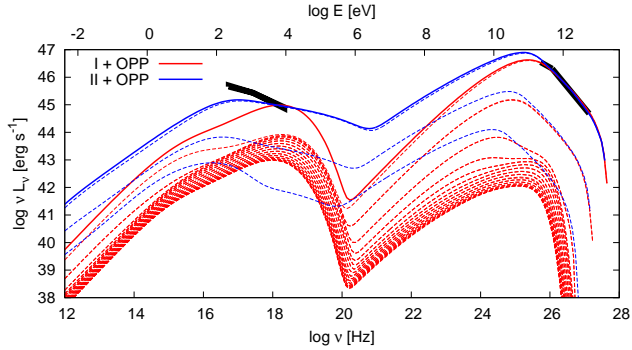}
  \caption{SEDs of a system of $N_{\rm ring}=30$ minijets evenly spaced around the jet axis, so that only one is directed close to the line-of-sight. Individual minijet spectra are shown with \emph{dashed lines}, summed spectrum with a \emph{solid line}. \emph{Red lines:} minijets calculated with model including Comptonization of radiation from the opposite minijet for case I. \emph{Blue lines:} minijets calculated for case II (only first 3 minijets and the sum off all 30 are shown for clarity).}
  \label{fig9}
\end{figure}

In Fig. \ref{fig9} we show observed luminosities of a series of $N_{\rm ring}=30$ minijets evenly spaced around the jet axis, forming a ring shown schematically in Fig. \ref{fig1}. Each minijet has exactly the same parameters and they differ solely by the observer's orientation. Only one minijet is closely aligned with the observer ($\theta\obs=0.1$; the highest line) and every consecutive one is rotated by an angle $2\pi/N_{\rm ring}=0.21$ in the $\mathcal{O}_1$ frame. In case I (\emph{red lines}), we find that even the second-best-aligned minijet (second highest line) is a negligible $\gamma$-ray emitter. However, in the hard X-ray band the ensemble of all misaligned minijets will contribute significantly. The summed spectrum of all individual minijets will be only marginally consistent with \emph{Chandra} limits. Therefore, the number of misaligned minijets per each aligned one should be less than 30 to explain the flare in PKS 2155-304. In case II (\emph{blue lines}), however, emission from misaligned minijets is negligible across the whole energy range. This is because the Lorentz factor in the island region is much higher for this model (`II+OPP`; $\Gamma_3=2.13$) than in model `I+OPP` ($\Gamma_3=1.06$).

\subsection{Lowering jet magnetization}

Jet magnetization of $\sigma_1\sim 100$ is required for reproducing fast TeV flares in the minijet model. It enables high minijet Lorentz factor $\Gamma_2$ and large average electron Lorentz factor $\left<\gamma_{e,2}\right>$, both of which are proportional to $\sqrt{\sigma_1}$. The latter requirement can be eased, noting that some sort of stochastic particle acceleration process is necessary in order to reproduce power-law spectral tails to fit observations. Lower $\sigma_1$ implies lower $\Gamma_2$ and thus a more compact minijet region for the same observed variability timescale. Reduced emitting volume would have a strong impact on the luminosity of both spectral components. We study this effect in the class of models including opposite minijet radiation, trying to match the TeV luminosity of PKS 2155-304 or, if impossible, calculating a model of maximum luminosity.

\begin{figure}
  \includegraphics[width=\columnwidth]{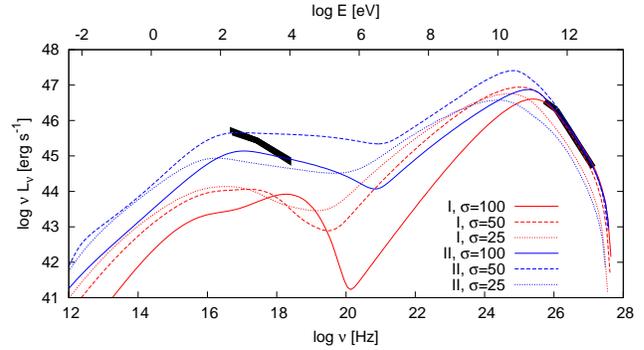}
  \caption{SEDs of minijets calculated for $\sigma_1=100$ (\emph{solid lines}; same as the dashed lines in Fig. \ref{fig8}), 50 (\emph{dashed lines}) and 25 (\emph{dotted lines}). The models have been calculated for cases of no (\emph{red}) or weak (\emph{blue}) guide field, including radiation from the opposite minijet.}
  \label{fig10}
\end{figure}

In Fig. \ref{fig10}, we compare the SEDs for three values of $\sigma_1$. The value $\sigma_1=100$ (\emph{solid lines}) has been used in \cite{2009MNRAS.395L..29G} and in previous paragraphs. $\sigma_1=50$ (\emph{dashed lines}) corresponds to $l_2=6.4\times 10^{14}\;{\rm cm}$ and $\Gamma_2=7.1$, while $\sigma_1=25$ (\emph{dotted lines}) to $l_2=4.5\times 10^{14}\;{\rm cm}$ and $\Gamma_2=5$. We were able to fit \emph{H.E.S.S.} data for PKS~2155-304 for $\sigma_1=50$, but not for $\sigma_1=25$, where opacity limits TeV luminosity below the observed level. Keeping $\Gamma\jet=10$, the last case corresponds to effective Lorentz factor of the minijet plasma $\Gamma_2\Gamma\jet\sim 50$, the minimum value derived by \cite{2008MNRAS.384L..19B}. Thus, our model confirms that prediction, even though it has been derived within a single-zone framework.

\section{Discussion}
\label{sec_disc}

Our calculations show that it is much easier to obtain a high Compton dominance for minijet models based on relativistic magnetic reconnection with no guide field (case I). Inspection of Table \ref{tab_param} reveals that this is related to two factors. First is a significantly lower magnetization of the minijet region plasma $\sigma_2$, which regulates the ratio of magnetic to electron pressure. For roughly the same thermal energy carried by particles in both cases, the magnetic energy density is 2 orders of magnitude lower in case I, and so is the synchrotron emissivity. The second reason is the much stronger compression of plasma crossing the stationary shock into the island region, leading to higher particle and magnetic pressure and thus higher synchrotron emission which is more strongly boosted back into the minijet region co-moving frame $\mathcal{O}_2$.

On the other hand, relativistic current sheets with no guide field have been found to develop a relativistic drift-kink instability (RDKI), which can disrupt the system before the particles can be non-thermally accelerated \citep{2008ApJ...677..530Z}. To explain TeV spectra in the flaring state of PKS 2155-304, a non-thermal power-law tail is needed in the electron distribution. A guide field has the effect of suppressing RDKI, allowing for efficient particle acceleration. Models with a significant guide field (case II) can satisfy observational constraints, when radiation by an opposite minijet is taken into account. In fact, this effect is much more pronounced in case II, increasing the Compton dominance by 2 orders of magnitude. Note, however, that these numerical studies were done for pair plasma, while in our model electron-proton plasma is required.

The amount of guide field in the minijets affects the spectrum in the soft X-ray band. This is independent of the slope of the non-thermal power-law tail (it is also true with no tail), but is related to the ratio between synchrotron components produced in the minijet and island regions. In case I, the spectrum is hard, because emission from the island region is stronger due to stronger plasma compression. In case II, the spectrum is soft, but still slightly harder than \emph{Chandra} spectrum of PKS~2155-304. In the flaring state of this object, a harder-when-brighter behaviour has been observed in both X-ray and TeV bands \citep{2009A&A...502..749A}. This can be understood if the brighter flares are produced by the unguided minijets, while the fainter flares (and some part of the quiescent emission) come from the guided ones. 

An isolated event like a TeV flare in PKS 2155-304 should be associated with a significant, brief and temporary change in jet physical parameters. A single disturbance comoving with the bulk jet flow would cover a distance $\Delta r\sim \Gamma\jet^2c\;\Delta t\sim 0.08(\Gamma\jet/10)^2(\Delta t/1\;{\rm d})\;{\rm pc}$. Thus, a $\sim 4$-day-long period of high activity would be related to a single global reversal of jet magnetic field travelling about $0.3\;{\rm pc}$. If such a flare were triggered by a factor external to the jet flow, it would produce a much longer activity period.

An alternative scenario for the origin of minijets would involve kink instabilities developing in jets dominated by toroidal magnetic fields. During a flare, the jet would experience a brief global instability. A physical mechanism triggering it would be related to some internal disturbance of the jet flow and not an external factor. Also, if the minijets arise from kink instabilities, the current sheet would have a more irregular structure and a case of two minijets aligned head-on is less likely. Thus, in this case it would be difficult to obtain a high Compton ratio, unless substantial external radiation fields are present in the reconnection region.

Observations of PKS 2155-304 indicate that emission in the quiescent state may be of different origin than the bulk of emission in the flaring state \citep{2010arXiv1005.3702H} and that what we observe in the flaring state is a superposition of these two components \citep{2009A&A...502..749A}. If the quiescent state emission is of the same level during the flaring state, it might illuminate the minijets, providing some external radiation. Assume that the source of this emission is located downstream from the minijets zone and has the same bulk Lorentz factor $\Gamma\jet=10$. The energy density of this external radiation of observed luminosity $L\obs$ in $\mathcal{O}_2$ is $u_{\rm ext}''\sim L\obs\Gamma_2^2/(32\pi\Gamma\jet^4 r^2c)$, where $r$ is the distance between the stationary radiation source and the current sheet in the frame external to the jet. For an observed X-ray luminosity in the quiescent state $L\obs\sim 3\times 10^{45}\;{\rm erg\;s^{-1}}$ \citep{2009ApJ...696L.150A} and $\Gamma_2=10$, we obtain $u_{\rm ext}''\sim 10\;r_{\rm 15}^{-2}\;{\rm erg\;cm^{-3}}$. This is comparable with values shown in the upper left panel of Fig. \ref{fig4}, given that $r\sim 10^{15}\;{\rm cm}$, which is extremely close in terms of typical minijet size $l_2\sim 10^{14-15}\;{\rm cm}$. It would be completely unimportant, if it arrived from a distance $r\sim 10^{18}\;{\rm cm}$, which is a typical location for the blazar zone in conventional models. Quiescent emission could be produced at very short distances from the cental black hole by minijets driven by kink instabilities.

There are now indications that $\gamma$-ray emission may be also anisotropic in bright blazars. \cite{2010A&A...512A..24S} calculated comoving-frame viewing angles of a large sample of blazars (mostly Flat-Spectrum Radio Quasars), based on VLBA estimates of both Lorentz and Doppler factors. They have found that \emph{Fermi/LAT}-bright sources fell into a range of viewing angles $40^\circ<\theta'<110^\circ$, while those sources not detected in 3 months of \emph{Fermi/LAT} monitoring had an almost uniform distribution in the full range of angles. This result cannot be easily explained without assuming internal anisotropy of the emitting region. The minijets model provides a physical structure that produces emission concentrated in the direction perpendicular to the jet bulk flow, as measured in the jet co-moving frame. On the other hand, variability constraints from \emph{Fermi/LAT} are not strong enough to discriminate this from other possibilities.

\section{Summary}
\label{sec_sum}

We have described a detailed model of minijets combining a dynamical solution of relativistic magnetic reconnection with a weak guide field with calculations of non-thermal radiative processes including evolution of the electron distribution, pair-production opacity and Compton drag with an exact treatment of the Klein-Nishina cross section. Here are our main results:

\begin{itemize}
\item
TeV luminosities produced in models using parameter values derived for a spherical blob model in \cite{2009MNRAS.395L..29G} are much lower than those observed during fast TeV flares in PKS 2155-304. This is because the minijet model presented here is located directly within the reconnection region of tiny opening angle.
\item
Maximum $\gamma$-ray luminosity obtained for models with maximum minijet region length allowed by observed variability timescale of 5 min is $\sim 10^{48}\;{\rm erg\;s^{-1}}$ both for models with no or weak guide field and regardless of whether one takes into account radiation from an opposite minijet system. The luminosity can be limited either by opacity or strong Compton drag decelerating the minijet flow. External radiation from the opposite minijet allows one to obtain the maximum luminosity with higher Compton dominance.
\item
SEDs matching the \emph{H.E.S.S.} spectral fits for the flaring state of PKS 2155-304 differ widely with respect to the X-ray flux level. Simultaneous \emph{Chandra} data are inconsistent with models including a significant guide field and excluding radiative interaction with an opposite minijet. In models with no guide field, radiation from misaligned minijets has to be taken into account in the X-ray band.
\item
Our model can be fitted to PKS 2155-304 data only for jet magnetization $\sigma_1>25$. Assuming bulk jet Lorentz factor $\Gamma\jet\sim 10$, this is consistent with one-zone results of \cite{2008MNRAS.384L..19B}. The minijet concept allows one to reconcile modest bulk jet Lorentz factors with large local Lorentz factors in the TeV flaring region.
\end{itemize}

\appendix

\section{Summary of radiative processes}
\label{app_rad}

Synchrotron emissivity is calculated from the formula \citep{1986A&A...164L..16C,2003A&A...406..855M}:
\be
\label{eq_em_syn}
j_{\rm SYN}(\nu)= \frac{3\sqrt{3}}{4\pi^2}\frac{\sigma_Tcu_B}{\nu_B}\int d\gamma\;n_e(\gamma)\mathcal{R}\left(\frac{\nu}{3\gamma^2\nu_B}\right)\,,
\ee
where $\mathcal{R}(x)=x^2[K_{1/3}(x)K_{4/3}(x)-0.6x({K_{4/3}(x)}^2-{K_{1/3}(x)}^2)]$, $u_B=B^2/(8\pi)$ and $\nu_B=eB/(2\pi m_ec)$. Since $\int\mathcal{R}(x)dx=4\sqrt{3}\pi/81$, one finds that the frequency-integrated formula is
\be
\label{eq_em_syn_ave}
j_{\rm SYN}=\frac{\sigma_Tc}{3\pi}n_eu_{\rm B}\left<\gamma_e^2\right>\,.
\ee

The IC radiation emissivity is calculated from \citep{1981Ap&SS..79..321A,2005MNRAS.363..954M}:
\bea
\label{eq_em_ic_mu_nu}
j_{\rm IC}(\nu,\mu) &=& \frac{3}{16\pi}\frac{h\sigma_T}{m_ec}\epsilon\int d\gamma\frac{n_e(\gamma)}{\gamma^2}\times\nonumber\\
&& \times\int d\epsilon_0\frac{u_0(\epsilon_0)}{\epsilon_0^2}f_\mu(w,b_\mu)\,,
\eea
where $\epsilon_0$ is the energy of the incident photon (in units of $m_ec^2$), $\epsilon$ is the energy of the scattered photon, $u_0(\epsilon_0)$ is the energy density spectrum of incident radiation and $\mu$ is the cosine of the scattering angle, $w=\epsilon/\gamma$, $b_\mu=2\epsilon_0\gamma(1-\mu)$. For the $f_\mu$ function see Eq. (\ref{eq_f}). The frequency-integrated formula is:
\bea
j_{\rm IC}(\mu) &=& \frac{\sigma_Tc}{4\pi}(1-\mu)^2\int d\gamma\,\gamma^2\,n_e(\gamma)\times\nonumber\\
&& \times\int d\epsilon_0\,u_0(\epsilon_0)f_{\rm KN,\mu}(b_\mu)\,,
\eea
where
\bea
f_{\rm KN,\mu}(b_\mu) &=& \frac{3}{b_\mu^2}\left[-\frac{5}{12}+\frac{11}{2b_\mu}+\frac{(3+b_\mu)(2+3b_\mu)}{12b_\mu(1+b_\mu)^3}+\right.\nonumber\\
&&\left.+\frac{(b_\mu-6)(b_\mu+2)}{2b_\mu^2}\ln(1+b_\mu)\right]\,.
\eea
In the Thomson limit ($b_\mu\ll 1$) $f_{\rm KN,\mu}(b_\mu)\sim 1$, thus
\be
\label{eq_em_ic_mu}
\frac{j_{\rm IC}(\mu)}{j_{\rm SYN}} \sim \frac{3}{4}(1-\mu)^2\frac{u_0}{u_{\rm B}}\,.
\ee
When the incident radiation field is isotropic, we can average the above formulae over the scattering angle:
\bea
\label{eq_em_ic_iso_nu}
j_{\rm IC}(\nu) &=& \frac{3}{16\pi}\frac{h\sigma_T}{m_ec}\epsilon\int d\gamma\frac{n_e(\gamma)}{\gamma^2}\int d\epsilon_0\frac{u_0(\epsilon_0)}{\epsilon_0^2}f(w,b)\,,\\
j_{\rm IC} &=& \frac{\sigma_Tc}{3\pi}\int d\gamma\,\gamma^2n_e(\gamma)\int d\epsilon_0\,u_0(\epsilon_0)f_{\rm KN}(b)\,,
\eea
where $b=4\epsilon_0\gamma$,
\bea
f(w,b) &=& 1+\frac{w^2}{2(1-w)}+\frac{w}{b(1-w)}-\frac{2w^2}{b^2(1-w)^2}+\nonumber\\
&&-\frac{w^3}{2b(1-w)^2}-\frac{2w}{b(1-w)}\ln\left[\frac{b(1-w)}{w}\right]\,\\
f_{\rm KN}(b) &=& \frac{9}{b^3}\left[-6-b+\frac{b^3}{12(1+b)^2}+\right.\nonumber\\
&& \left.+\left(\frac{b}{2}+6+\frac{6}{b}\right)\ln(1+b)+2{\rm Li}_2(-b)\right]\,.
\eea
In the Thomson limit, $j_{\rm IC}/j_{\rm SYN} \sim u_0/u_{\rm B}$.

Radiative cooling of electrons is calculated from \citep{2003A&A...406..855M,2005MNRAS.363..954M}:
\bea
\label{eq_cool}
\left.\der{\gamma}{x}\right|_{\rm rad} &=& -\frac{4\sigma_T}{3m_ec^2}\frac{(\gamma^2-1)}{\sqrt{\Gamma^2-1}}\times\nonumber\\
&&\times\left[u_B+\int d\epsilon_0u_0(\epsilon_0)f_{\rm KN}(b)\right]\,.
\eea

The pair-production absorption coefficient from a directed beam of ambient photons of energy $\epsilon_0=h\nu_0/(m_ec^2)$ and energy density $u_0$ is \citep{1967PhRv..155.1404G}
\be
\label{eq_kappa_pp}
\kappa_{\gamma\gamma}(\epsilon,\mu) = \frac{1}{m_ec^2}\int d\epsilon_0\frac{u_0(\epsilon_0)}{\epsilon_0}\,\sigma_{\gamma\gamma}{\left(\frac{1-\mu}{2}\epsilon\,\epsilon_0\right)}\,,
\ee
where
\bea
\label{eq_sigma_pp}
\sigma_{\gamma\gamma}(x) &=& \frac{3}{16}\sigma_T\left(1-\beta^2\right)\times\nonumber\\
&&\times\left[\left(3-\beta^4\right)\ln{\left(\frac{1+\beta}{1-\beta}\right)}-2\beta\left(2-\beta^2\right)\right]
\eea
and $\beta=\sqrt{1-1/x}$.

\section{Compton drag in Klein-Nishina regime}
\label{app_drag}

An integral formula for radiative force from the IC process taking into account the Klein-Nishina cross-section was introduced in \cite{1974ApJ...188..121B} and used for studying Compton drag in \cite{2000ApJ...534..239M}. We provide here an analytical formula, which is valid for an isotropic distribution of relativistic electrons of random Lorentz factor $\gamma\gg 1$, implying $\epsilon\gg\epsilon_0$, where $\epsilon_0$ is the energy of incident photons (in units of $m_ec^2$) and $\epsilon$ is the energy of scattered photons. In this approximation, the differential IC cross-section is (Eq. 20 in \citealt{1981Ap&SS..79..321A}; see also \citealt{2005MNRAS.363..954M}):
\be
\frac{d\sigma_{\rm IC}}{d\epsilon\;d\Omega}(\epsilon_0,\gamma)=\frac{3\sigma_T}{16\pi\gamma^2\epsilon_0}f_\mu(w,b_\mu)\,,
\ee
where $\epsilon$ is the energy of scattered photons, $\mu$ is the cosine of the scattering angle, $w=\epsilon/\gamma$, $b_\mu=2\epsilon_0\gamma(1-\mu)$,
\be
\label{eq_f}
f_\mu(w,b_\mu) = 1+\frac{w^2}{2(1-w)}-\frac{2w}{b_\mu(1-w)}+\frac{2w^2}{b_\mu^2(1-w)^2}\,.
\ee
The energy of scattered photons is limited to $\epsilon\le \gamma b_\mu/(1+b_\mu)$.

For a single scattering the electron loses energy of $\sim \epsilon m_ec^2$ and gains momentum $\sim-\mu\epsilon m_ec$ in the direction of the incident photon. We calculate the average force per electron exerted by a directed photon beam of number density $n_{\rm ph}(\epsilon_0)$:
\bea
\dot{p}_e(\gamma) &\sim& -m_ec^2 \int d\epsilon_0\;n_{\rm ph}(\epsilon_0)\int d\epsilon\;\epsilon\int d\Omega\;\mu\frac{d\sigma_{\rm IC}}{d\epsilon\;d\Omega}(\epsilon_0,\gamma)\nonumber\\
&\sim& \frac{2}{3}\sigma_T\gamma^2\int d\epsilon_0\;u_{\rm ph}(\epsilon_0)\;f_{\rm KN,1}(b)\,,
\label{eq_drag_force_1e}
\eea
where $u_{\rm ph}(\epsilon_0)=\epsilon_0m_ec^2\;n_{\rm ph}(\epsilon_0)$,
\bea
f_{\rm KN,1}(b) &=& -\frac{9}{\gamma^2b^2}\int_0^{\epsilon_{\rm max}} d\epsilon\;\epsilon\int_{-1}^{\mu_{\rm max}} d\mu\;\mu\;f_\mu(w,b) = \nonumber\\
&=&\frac{18}{b^3}\left[22+\frac{(2+3b)b^2}{12(1+b)^2}-10\frac{1+b}{b}\ln(1+b)+\right.\nonumber\\
&&\left.+2\frac{6-b}{b}{\rm Li}_2(-b)\right]\,,
\eea
$\mu_{\rm max}=1-w/[2\epsilon_0\gamma(1-w)]$, $\epsilon_{\rm max}=\gamma b/(1+b)$, $b=4\epsilon_0\gamma$ and ${\rm Li}_2$ is the dilogarithm function.

In the Thomson limit ($b\ll 1$), it can be shown that $f_{\rm KN,1}(b)\sim 1$ and so Eq. (\ref{eq_drag_force_1e}) is consistent with Eq. (8) in \cite{1996MNRAS.280..781S}.

\section{Radiation energy density from a stationary pattern of relativistic flow}
\label{app_urad}

A recurring problem in this work is to calculate the energy density of synchrotron and external radiation in a frame co-moving with a relativistic flow. The emitting region is stationary in frame $\mathcal{O}$ and the emitting fluid is characterized by bulk Lorentz factor $\Gamma_{\rm em}$ and co-moving emissivity $j'(\nu_{\rm em}')$. Consider a small element of the emitting region of volume $\Delta V$, observed from distance $r$, with fluid velocity making angle $\theta$ with the direction away from the observer. At the observer's position, there is a flow of Lorentz factor $\Gamma\obs$ and velocity parallel to the velocity of the emitting fluid. In $\mathcal{O}$, the radiation energy density is $\Delta u(\nu)=j(\nu)\Delta V/(r^2c)$. Emissivity in $\mathcal{O}$ is given by $j(\nu)=\mathcal{D}_{\rm em}^2 j'(\mathcal{D}_{\rm em}^{-1}\nu)$, where $\mathcal{D}_{\rm em}=[\Gamma_{\rm em}(1+\beta_{\rm em}\cos\theta)]^{-1}$. A transformation of energy density to the fluid frame at observer's position is $\Delta u'(\nu')=\mathcal{D}_{\rm obs}' \Delta u({\mathcal{D}_{\rm obs}'}^{-1}\nu')$, where $\mathcal{D}_{\rm obs}'=\Gamma_{\rm obs}(1+\beta_{\rm obs}\cos\theta)$. In effect, we have obtained:
\be
\label{eq_urad_patt1}
\Delta u'(\nu') = \mathcal{D}_{\rm obs}'\mathcal{D}_{\rm em}^2\frac{j'\left({\mathcal{D}_{\rm obs}'}^{-1}\mathcal{D}_{\rm em}^{-1}\nu'\right)}{c}\frac{\Delta V}{r^2}\,.
\ee

In application to radiation energy density in the minijet co-moving frame, we would take $\Gamma_{\rm obs}=\Gamma_2$ and $\Gamma_{\rm em}=\Gamma_3$ for radiation emitted in the island region or $\Gamma_{\rm em}=\Gamma_2$ for synchrotron radiation emitted in the minijet region. In the case of $\Gamma_{\rm em}=\Gamma_{\rm obs}$, formula (\ref{eq_urad_patt1}) reduces to
\be
\label{eq_urad_patt2}
\Delta u'(\nu') = \mathcal{D}_{\rm em}\frac{j'\left(\nu'\right)}{c}\frac{\Delta V}{r^2}\,.
\ee

\section*{Acknowledgments}

KN thanks the JILA staff for their hospitality and support during his visit. KN and MS have been supported by the Polish Ministry of Science and Higher Education grants N~N203 301635 and N~N203 386337, and by the Polish Astroparticle Network grant 621/E-78/SN-0068/2007. DG acknowledges support from the Lyman Spitzer, Jr. Fellowship awarded by the Department of Astrophysical Sciences at Princeton University. MCB acknowledges support from NASA through the Fermi Guest Investigator and Astrophysics Theory Programs, and from NSF through grant AST-0907872.

\end{document}